\documentclass[12pt]{iopart}
\usepackage{iopams}
\usepackage{graphicx}
\usepackage[colorlinks=true,citecolor=blue,linkcolor=magenta]{hyperref}
\usepackage{bm}
\usepackage{dsfont}
\usepackage{color,psfrag}
\usepackage{pstool}
\usepackage{array}
\usepackage{longtable}
\usepackage{wrapfig}
\usepackage{bbm}
\newcommand{\bra}[1]{\langle #1|}
\newcommand{\ket}[1]{|#1\rangle}

\begin{document}

\title{Coupling of nitrogen vacancy centers in nanodiamonds by means of phonons}

\author{A Albrecht$^{1,2}$, A Retzker$^{3}$, F Jelezko$^{4,2}$ and M B Plenio$^{1,2}$}
\address{$ˆ1$ Institut f\"ur Theoretische Physik, Albert-Einstein-Allee 11, Universit\"at Ulm, 89069 Ulm, Germany} 
\address{$ˆ2$ Center for Integrated Quantum Science and Technology, Universit\"at Ulm, 89069 Ulm, Germany} 
\address{$ˆ3$ Racah Institute of Physics, The Hebrew University of Jerusalem, Jerusalem 91904, Israel} 
\address{$ˆ4$ Institut f\"ur Quantenoptik, Albert-Einstein-Allee 11, Universit\"at Ulm, 89069 Ulm, Germany}
\ead{andreas.albrecht@uni-ulm.de}

\begin{abstract}
Realising controlled quantum dynamics via the magnetic interactions between colour centers in diamond remains a challenge despite recent demonstrations for nanometer separated pairs.
Here we propose to use the intrinsic acoustical phonons in diamond as a data bus for accomplishing this task. We show that for nanodiamonds the electron-phonon coupling can take significant values that together with mode frequencies in the THz range, can serve as a resource for conditional gate operations. Based on these results we analyze how to use this phonon-induced interaction for constructing quantum gates among the electron-spin triplet ground states, introducing the phonon dependence via Raman transitions. Combined with decoupling pulses this offers the possibility for creating entangled states within nanodiamonds on the scale of several tens of nanometers, a promising prerequisite for quantum sensing applications.
\end{abstract}

\maketitle

\section{Introduction}
Tremendous progress in understanding and manipulating the nitrogen vacancy (NV) center in diamond throughout the last decade revealed its promising capabilities for quantum information and sensing applications. The basic foundations of coherent manipulation, high fidelity polarization and optical readout\,\cite{jelezko04} paved the way for using the NV center as a fully controllable quantum bit capable of operating at room temperature with extraordinary long coherence times that may reach the millisecond range\,\cite{balasubramanian09}. Whereas the coherent coupling and entanglement to nuclear spins of nitrogen\,\cite{gaebel06, vanderSar12} and carbon-13\,\cite{dutt07, neumann08} has been demonstrated in numerous experiments, bringing different nitrogen vacancy centers to interaction remains challenging and has been demonstrated only recently\,\cite{dolde13, bernien12}. One approach for coupling distinct NV-centers makes use of their dipolar interactions\,\cite{dolde13, albrecht13, bermudez11}, which is limited by the strong distance dependence of the coupling and therefore has been demonstrated only for very closely separated pairs. Another method consists of interconnecting the NV center solid state spin qubits with photons\,\cite{bernien12}, that has lead to extensive research in the design of cavities and photon couplings\,\cite{wolters10, riedrich11, faraon12, hausmann12}. In contrast to that, the coupling to phonons is much less studied. Whereas this mechanism serves as the prominent data bus for conditional quantum operations in the trapped ion approach to quantum computing\,\cite{leibfried2003} and has been proposed to allow even for a strong coupling regime in phonon cavity structures in silicon\,\cite{soykal11}, intrinsic phonon coupling is assumed to be inaccessible in macroscopic diamonds at room temperature.  However, the  coupling to magnetized nanomechanical oscillators as AFM cantilevers  was successfully performed, allowing for the sensing of the vibrational mode\,\cite{kolkowitz12, arcizet11} and even for the coherent manipulation of the NV center electron spin state\,\cite{hong12}, that might provide the basic ingredient for future phonon mediated quantum networks\,\cite{rabl10}. 
Here we will show that a significant intrinsic phonon coupling can be expected in nanodiamonds at low temperatures. Those can be fabricated down to 4\,nm in size, additionally being capable of hosting NV centers\,\cite{tisler09}. We study the coupling strength to long wavelength (low frequency) acoustical modes and analyze the possibility to exploit these global modes for entanglement operations by creating a Raman-induced phonon coupling within the ground state electron triplet states of the NV center. 

\section{Coupling of the nitrogen vacancy center to phonons}\label{sect_phon_coupl}
\begin{figure}[htb]
\begin{centering}
\includegraphics[scale=0.30]{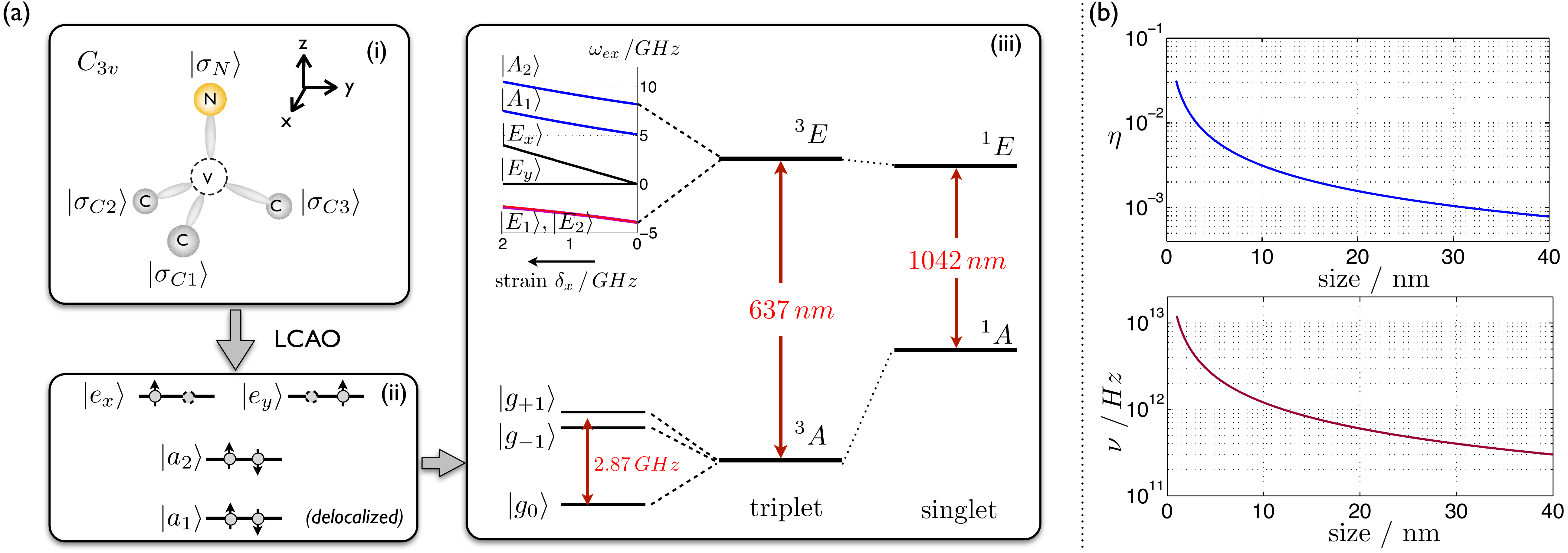}
\caption{\label{b_levelstrain} \textbf{Energy level structure and phonon coupling} (a) NV center energy levels and composition out of dangling bond orbitals. (i) NV center and dangling bond orbitals with symmetry $C_{3v}$. (ii) Symmetric linear combinations of the dangling bond orbitals (LCAO) lead to the electronic states $\ket{a_1}, \ket{a_2}, \ket{e_x}, \ket{e_y}$, occupied with the six electrons (or equivalently two holes) of the NV center. (iii) Combining the electronic and spin wavefunction leads to the NV center energy level structure with spin singlets and triplets. Ground states correspond to two holes in the $\ket{e_{x/y}}$ state as illustrated in (ii) and excited states to one hole in $\ket{a_2}$ and $\ket{e_{x/y}}$. Strain and therefore phonons cause displacements of the dangling orbitals (i), and consequently modify the energy levels in (ii) and (iii). The influence of the strain $\delta_x=-\zeta\,\epsilon_{xx}$ on the excited state fine structure is illustrated in the inset. \textbf{(b)} Phonon coupling coefficient $\eta$ (\ref{etaform}) and frequency\,$\nu$ of the lowest acoustical mode vs size (diameter) of a spherical nanodiamond.}
\end{centering}
\end{figure}

The localization of the NV-center electronic states to the vacancy defect itself permits their description as superpositions of molecular orbitals, each associated with a dangling bond orbital of the atoms involved in the defect center\,\cite{maze11, doherty11}. That is, according to the $C_{3v}$ symmetry of the defect center, four electronic states $\mathcal{M}_{\mathrm{el}}=\{ \ket{a_1}, \ket{a_2}, \ket{e_x}, \ket{e_y}  \}$  can be constructed out of the dangling bond orbitals $\mathcal{M}_{\mathrm{db}}=\{\ket{\sigma_C^1}, \ket{\sigma_C^2}, \ket{\sigma_C^3}, \ket{\sigma_N}  \}$ by linear combinations, with the index $C$ referring to carbon and $N$ to the nitrogen related bonds (see figure\,\ref{b_levelstrain} (a)).

   In this framework the well known ground and excited state energy level structure of the $\mathrm{NV}^-$- center follows by associating the six electrons involved with the available electronic states $\mathcal{M}_{\mathrm{el}}$, taking the additional spin properties into account. Equivalently, and more simply, this can be described by two holes relative to a complete filling\,\cite{maze11}. Calculating the coupling to phonons can be decomposed into two steps by noting that the long wavelength acoustical phonons considered here will introduce a periodic strain to the NV center. Thus in a first step we will discuss the effect of strain to the energy level structure and in a second step link the vibrational phonon mode to the strain property. For the former case we will follow the discussion presented in\,\cite{maze11} (see also a similar discussion in\,\cite{doherty11}). The electron-nuclei Coulomb coupling can be described by an interaction of the form
\begin{equation}\label{scoulomb1}  V=\sum_{i}g_{i}\,\ket{\sigma_i}\bra{\sigma_i}+\sum_{i>j} h_{ij}(|\vec{r}_{ij}|)\,\ket{\sigma_{i}}\bra{\sigma_{j}}+\mathrm{h.c.} \end{equation}  
with $i$ and $j$ describing a summation over all possible $\mathcal{M}_{\mathrm{db}}$ configurations and $\vec{r}_{ij}=\vec{r}_i-\vec{r}_j$ with $\vec{r}_{i/j}$ the corresponding atom position vectors. Noting that the coupling coefficient depends on the relative distance of the atoms involved, it is obvious that stress related displacements $\vec{u}_i$ will influence the energy level structure and for small displacements, as expected by phonon effects,  the strain perturbation on the level of dangling bonds can be described by
\begin{equation}\label{scoulomb2}  \delta V=\sum_{i>j}\delta h_{ij}(|\vec{r}_{ij}|)\,\, \ket{\sigma_i}\bra{\sigma_j} +\mathrm{h.c.}   \end{equation}
with $\delta h_{ij}=\nabla_{\vec{r}}h_{ij}\,\delta\vec{u}=\nabla_{\vec{r}}h_{ij}\,\mathbf{e}\vec{r}$ the first order correction following from an expansion of $h_{ij}$. Herein the tensor $\mathbf{e}_{\mu \nu}=\partial u_\mu/(\partial r_\nu)$ is related to the strain tensor $\mathbf{\epsilon}$ by $\mathbf{\epsilon}_{\mu \nu}=1/2(\mathbf{e}_{\mu \nu}+\mathbf{e}_{\nu \mu})$. Considering a specific ground state to excited state triplet transition this results in the following Hamiltonian for the NV-center energy levels:
\begin{equation}\label{dhstrain}  H_{\mathrm{el-phon}}\simeq 2\,\delta_1\,|g\rangle\langle g|+|e\rangle\langle e| (\delta_1+\delta_4+\xi) \end{equation}
wherein $\delta_1=-\zeta (\epsilon_{xx}+\epsilon_{yy})$ and $\delta_4=\zeta 8\,\beta^2\,e_{zz}$ with\,\cite{maze11} $\zeta\simeq 610\,THz$ and $\beta$ accounting for the difference of the nitrogen coupling compared to the carbon related ones. $\ket{g}$ can be any of the triplet ground states, that exhibit an equal shift under the influence of strain, whereas for the excited states we find $\ket{e}$ $\xi=\pm \zeta\,\sqrt{4\,\epsilon_{xy}^2+[\epsilon_{xx}-\epsilon_{yy}]^2}  $ for the states $\ket{E_x}$ and $\ket{E_y}$, respectively, and $\xi=0$ otherwise. For more details on the derivation of\,(\ref{dhstrain}) we refer to\,\ref{append_phon_coupl}.\par
Vibrational modes will introduce a time-dependent periodic strain to the system, that exhibits a simple description for long wavelength acoustical modes with wavevectors near the Brillouin-zone. In that case, and assuming periodic boundary conditions, the discrete Bloch-type wavefunction, describing the displacement of the crystal lattice positions, can be described by a continuous displacement field in space $\vec{u}(\vec{r})$\,\,\cite{ashcroft05book} replacing the discrete lattice positions $n$ by $\vec{r}_n\to r$ and the displacement for a specific mode $\alpha$ is thus obtained by\,\cite{yu99book}
\begin{equation}\label{modedispl} \mathbf{e}_{ij}=\sqrt{\frac{\hbar}{2 M\,\nu(k)}}\,k_j\,e_i^{(\alpha)}\,(-i)\,\left(a_{\alpha}^\dagger-a_{\alpha} \right)  \end{equation}
with $\vec{k}$ the wavevector, $\vec{e}^{\,(\alpha)}$ the mode eigenvector, $\nu(k)$ the angular frequency and $a_\alpha, a_\alpha^\dagger$ the phonon creators and annihilators of the mode, respectively,  and $M$ the mass of the nanodiamond. For the linear acoustical modes near the zone center $\nu(k)=c\cdot k$ with $c=1.2\cdot 10^4\, m/s$ the speed of sound in diamond and $k=2\,\pi/l$ out of applying periodic boundary conditions, with $l$ the diamond length in the corresponding mode direction. Combining equations (\ref{dhstrain}) and (\ref{modedispl})  allows for the calculation of the energy shift associated with the phonon coupling and exhibits the typical form of a deformation potential coupling describing local lattice compression and dilation as expected for the long wavelength acoustical phonon case. As an illustration let us explicitly give the electron-phonon coupling energy shift Hamiltonian for the specific case of a mode $\alpha$ with $\vec{e}^{\,(\alpha)}\parallel \vec{k}$ pointing in x-direction and choosing $\ket{e}=\ket{A_2}$ as the excited state, favourable in the sense that it is not coupled to the singlet state via spin-orbit coupling\,\cite{maze11, togan11}. In that case the electron-phonon coupling takes the simple form
\begin{equation}  H_{\mathrm{el-phon}}=-\eta\,\nu\,(-i)\,(a^\dagger-a)\,\ket{e}\bra{e}  \end{equation}
with the coupling coefficient
\begin{equation}\label{etaform}  \eta=\zeta\,\frac{k}{\nu}\sqrt{\frac{\hbar}{2\,M\,\nu}} \, \end{equation}
that takes the role of the well-known Lamb-Dicke parameter.\par
Note that the periodic boundary condition treatment is strictly valid only for the case of an infinite crystal; for a finite nanodiamond crystal, shape-dependent confinement effects alter the boundary conditions, therefore leading to a modification of the specific phonon mode spectrum. We will discuss the NV-center phonon coupling in that case for the example of a free elastic sphere in\,\ref{append_elast_sphere}. Although a change in the exact microscopic behaviour is observed in that case, the coupling properties of the long-wavelength radial breathing mode agree fairly well with the periodic boundary treatment. Moreover the scaling properties remain unchanged for both descriptions. 

Adding a laser coupling with Rabi frequency $\Omega$ and frequency $\omega_L$ to couple one of the ground states to the excited state $\ket{e}$ and applying the canonical Schrieffer-Wolff transformation $U=\mathrm{exp}(i\eta\, (a^\dagger+a)\ket{e}\bra{e})$ beside restricting the discussion to a single mode, leads to the following Hamiltonian in the rotating wave approximation\,\cite{wilson04, mintert01} (a discussion on how to include the mode relaxation and the finite excited state lifetime can be found in \ref{append_decay})
\begin{equation}\label{hamcoupl1}  H=\frac{\tilde{\omega}_0}{2}\,\sigma_z+\nu a^\dagger a+\left[ \frac{\Omega}{2}\,\ket{e}\bra{g}\,e^{-i\omega_L t}\,e^{i\,\eta (a^\dagger+a)}+\mathrm{h.c.}  \right]\,. \end{equation}
Herein $\sigma_z$ refers to the Pauli operator of the corresponding transition and $\tilde{\omega}_0=\omega_0+\eta^2\,\nu$ with $\omega_0$ the transition frequency and the second part following from the Schrieffer-Wolff transformation. Note also that the phonon coupling mechanism is assumed to be completely originated from the electron-phonon coupling Hamiltonian of the corresponding transition, whereas the photon recoil based coupling mechanism has been neglected as it is orders of magnitude lower in macroscopic systems\,\cite{wilson04}.  The coupling parameter $\eta$\,(\ref{etaform}) for different nanodiamond sizes is shown in figure\,\ref{b_levelstrain}\,(b) taking typical values of the order of $10^{-2}-10^{-3}$ and decreasing with the nanodiamond radius $\propto R^{-1}$ for a spherical diamond of radius $R$. For general diamond shapes $\eta\propto \sqrt{l/V}$ with $l$ the length in the mode direction and $V$ the diamond volume, leading e.g. to a $\eta\propto 1/\sqrt{R}$ scaling for a two-dimensional structure of area $\propto R^2$. The mode frequencies $\nu$ for the lowest energy mode, for a spherical diamond also representing the frequency difference among neighbouring modes, is shown in figure\,\ref{b_levelstrain}\,(b) taking typical values in the THz-range. These high mode frequencies are advantageous in the sense that neighbouring modes are well separated and additionally thermal occupation probabilities are low. The decreasing magnitude of both the coupling strength and the mode frequency limits the diamond size to several tens of nanometers.   Hamiltonian\,(\ref{hamcoupl1}) can be expanded in orders of the small coupling parameter $\eta$ with the zeroth phonon independent order providing the carrier and the first order the blue ($\propto a^\dagger |e\rangle\langle g| $) and red sideband ($\propto a|e\rangle\langle g|$) transitions.\\
These phonon sidebands should be experimentally observable in low temperature emission spectra\,\cite{cirac93}; however no such experimental investigations with nanodiamonds $<5\,nm$ have been performed to date.

\section{Phonon mediated gate interaction } 
\begin{figure}[htb]
\begin{centering}
\includegraphics[scale=0.30]{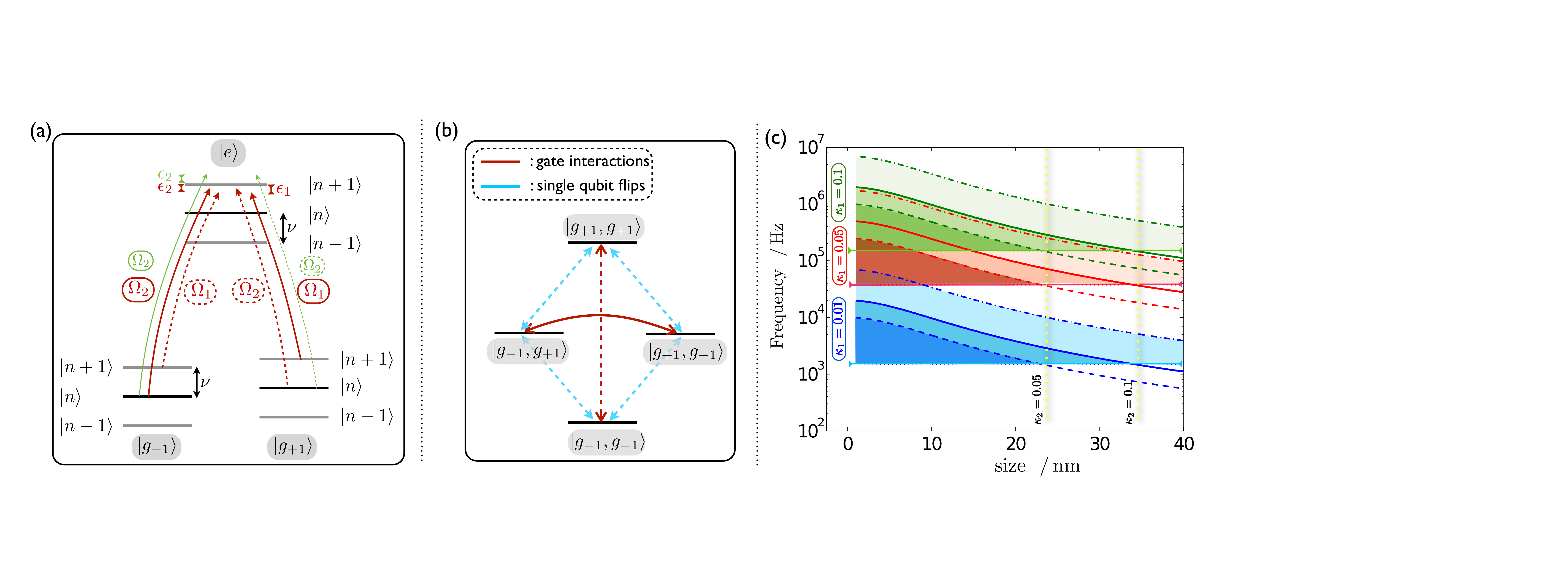}
\caption{\label{b_gatesetup} \textbf{Phonon mediated gate} \textbf{(a)} Setup for creating a Raman induced phonon coupling within the ground state manifold by off-resonantly exciting the excited state with a carrier and blue-sideband transition. Dashed lines represent couplings that just appear in the double-path setup. Green lines are compensation pulses that become relevant for gates in the non-perturbative limit $\kappa_2=\tilde{\Omega}/\Delta\epsilon\lesssim 1$. \textbf{(b)} Final effective two-qubit gate interaction for the single (only continuous lines) and both-path (all interactions) configuration. \textbf{(c)} Effective gate frequency vs size (=diameter) of a spherical nanodiamond and different values of $\kappa_1$ and $\kappa_2$, taking the coupling to the second neighbouring excited state into account. Herein $\Omega_2=\kappa_1\,\nu$, $\Omega_1=\eta\,\Omega_2$ and $\kappa_2$ defined as in the text. Continuous, dashed and dashed dotted lines correspond to $\kappa_2=0.1, 0.05, 0.35$, whereas blue, red and green corresponds to $\kappa_1=0.01, 0.05, 0.1$, respectively. Horizontal lines (with triangle endpoints) denote the corresponding $\Gamma_{\mathrm{eff}}$ value that imposes a limit on the maximal nanodiamond size shown for $\kappa_2=0.05$ and $\kappa_2=0.1$ by the vertical yellow lines. The area for which the ratio $\Omega_{\mathrm{gate}}>\Gamma_{\mathrm{eff}}$ is coloured with decreasing saturation for increasing $\kappa_2$. Note that this ratio is independent of the choice of $\kappa_1$ for a given nanodiamond size.  }
\end{centering}
\end{figure}

The phonon dependent sideband transitions offer the possibility of correlating the NV center state with the global vibrational one. This can be used as a source for gate operations\,\cite{sorensen99, bermudez12, lemmer13} between different NV centers within the same nanodiamond. In one of its simplest versions a M\o lmer-S\o rensen type gate\,\cite{sorensen99, roos08} can be directly implemented on the ground-excited state transition, that has the advantage of being insensitive to the actual phonon state, therefore allowing gate operations even for thermal initial phonon states and circumventing the need for vibrational mode cooling. However such a concept suffers from the relatively short lifetime of the excited state limited by both optical decay to the ground state as well as non-radiative spin-orbit relaxations to singlet states. Moreover, in order to avoid off-resonant excitations to several excited states, the driving field strength has to be limited to values smaller than the typical energy gaps $\sim 4\, GHz$, therefore limiting the maximal gate speed.\par
Here we provide an alternative concept that allows to perform the gate within the triplet ground state manifold.  A Raman transition via the excited state provides the required phonon coupling, that otherwise would not exist according to the absence of a difference in the electron-phonon coupling among the ground state triplet states in\,(\ref{dhstrain}). For this setup to work, a $\Lambda$-transition between ground and excited states is required, that for the NV center exists between the $\ket{g_{+1}}, \ket{g_{-1}}$ electron-spin triplet ground states and either of the excited states ${\ket{A_1}, \ket{A_2}, \ket{E_1}, \ket{E_2}}$, the latter forming equal superpositions of the $m_s=\pm 1$ electron spin projections\,\cite{maze11}. These $\Lambda$-transitions have already been successfully implemented and analyzed in experiments\,\cite{togan10, togan11}, being accessible in the low temperature ($<10\,K$) and strain limit. Using circular polarized light allows to drive spin-selective transitions between those states that might be advantageous for tuning the detunings and couplings. Additionally the Raman-transition within the $\Lambda$-scheme has to be carefully tuned to a single sideband, that is either the red or blue one, as otherwise the phonon dependence of the effective ground state transition is cancelled by interference of those two paths. For this task the high phonon frequencies of nanodiamonds are advantageous as they allow the single sideband addressing without stringent conditions on the Rabi-frequency of the coupling field. The gate interaction itself follows from a two-step process: First, the Raman transition provides a phonon dependent coupling within the ground state manifold. Second, this phonon dependent coupling can be used to implement a gate between two NV-centers by off-resonantly exciting the phonon state similar to the direct M\o lmer-S\o rensen approach. That way the excited state relaxation is suppressed by the off-resonance of the Raman transition, leading to an improved ratio between gate and relaxation time compared to a direct gate implementation on the ground-excited state transition as will be discussed below. Moreover, contributions of different excited states simply add up, therefore allowing for a straightforward integration of this effect into the formalism and not leading to an excitation of several states as might happen in the direct implementation.

\subsection{Setup and first effective form} 
The setup for this gate is illustrated in figure\,\ref{b_gatesetup} (a) where the dashed transitions are present only in the double-path (dp) setup, that leads to a complete $\sigma_x\otimes \sigma_x$-type coupling in the ground state manifold, whereas they are absent in the single-path (sp) one resulting in a gate in the reduced manifold $\mathcal{M}_1=\{\ket{g_{+1},g_{-1}}, \ket{g_{-1},g_{+1}}  \}$. The corresponding Hamiltonian in a frame rotating with the laser frequency for NV center $k$ can be written as
\begin{eqnarray}\label{hrotframe} \fl H_{k}^{\mathrm{sp}}=&\frac{\Omega_1}{2}\,\ket{e}\bra{g_{+1}} \,e^{-i\epsilon_1\,t}+  \frac{\Omega_2}{2}\,\ket{e}\bra{g_{-1}}\,e^{-i\,(\nu+\epsilon_2)\,t}
&+i\,\eta_k\,\frac{\Omega_2}{2}\,a^\dagger\,\ket{e}\bra{g_{-1}}\,e^{-i\,\epsilon_2\,t}+\mathrm{h.c.}
\end{eqnarray} 
with $\Omega_1\simeq \eta_k\,\Omega_2$. Herein the first and last contribution form the phonon-dependent Raman transition consisting of a carrier and blue sideband excitation, respectively, provided that $\epsilon_1\gg \Omega_1$ and $\epsilon_2\gg \Omega_2$ (the same can be achieved with a red sideband interaction as well). The second contribution describes the unavoidable carrier excitation associated with the sideband term and is not required for the gate interaction itself, however cannot be neglected either. Higher order terms in $\eta_k$ have been omitted. For the double-path setup $H_{k}^{\mathrm{dp}}=H_{k}^{\mathrm{sp}}+H_{k}^{\mathrm{sp}}\bigl |_{\ket{g_{+1}}\leftrightarrow \ket{g_{-1}}}$ with the second contribution corresponding to the first one by replacing $\ket{g_{+1}}$ by $\ket{g_{-1}}$ and vice versa, i.e. both of the couplings $\Omega_1$ and $\Omega_2$ are present simultaneously on both transitions. Describing the off-resonance ratio between driving fields and detuning by $\kappa_1\ll 1$ ($\kappa_1=\Omega_1/\epsilon_1\simeq \eta\Omega_2/\epsilon_2$) and noting that $\epsilon_k\ll \nu$, the optimal choice of parameters is given by $\Omega_2\sim\kappa_1\,\nu$ and $\epsilon_1\simeq \epsilon_2=1/\kappa_1\,\eta\,\Omega_2$.  \\
Eliminating the off-resonant excited state  results in the effective Hamiltonians
\numparts
\begin{eqnarray}\label{heff1} H_{\mathrm{eff},k}^{\mathrm{I,dp}}&=\frac{\delta_k^{\mathrm{dp}}}{2}\,\sigma_x+\left[i\,\frac{\tilde{\Omega}_k}{2}\,e^{i\,\Delta\epsilon\,t}\,a^\dagger\left(\sigma_x+\mathds{1}\right)+\mathrm{h.c.}\right] \\
\label{heff1b} H_{\mathrm{eff},k}^{\mathrm{I,sp}}&=\frac{\delta_k^{\mathrm{sp}}}{2}\,\sigma_z+\left[i\,\frac{\tilde{\Omega}_k}{2}\,e^{i\,\Delta\epsilon\,t}\,a^\dagger\sigma_++\mathrm{h.c.}\right]
 \end{eqnarray}
\endnumparts
with $\tilde{\Omega}_k=\frac{1}{4}\,\Omega_1\,(\eta_k\,\Omega_2)\,(\epsilon_1+\epsilon_2)/(\epsilon_1\,\epsilon_2)$ and $\delta^{\mathrm{dp}}=\delta_{\mu=2}$ and $\delta^{\mathrm{sp}}=1/2\,\delta_{\mu=1}$ with $\delta_\mu=1/2\,\left( \Omega_1^2/\epsilon_1+(-1)^{\mu}\eta_k^2\,(1+\hat{n})\,\Omega_2^2 /\epsilon_2+(-1)^\mu\,\Omega_2^2/(\nu+\epsilon_2) \right)$. Moreover $\Delta\epsilon=\epsilon_1-\epsilon_2+\chi\, \eta_k^2\Omega_2^2/\epsilon_2 $ where the last term accounts for the coupling induced shift of the mode frequency with $\chi=1/4$ for the double and $\chi=1/8$ for the single-path configuration. The Pauli operators $\sigma_x$, $\sigma_z$ and $\sigma_+$ are defined in the ground state manifold $\{ \ket{g_{+1}}, \ket{g_{-1}} \}$ and $\hat{n}$ denotes the phonon number operator.  The terms involving phonon excitations correspond to paths involving both the sideband and carrier transition whereas non-combined paths lead to single transitions or AC Stark shifts not associated with an effective phonon (de-)\,excitation, respectively. Note that the last contribution in $\delta_\mu$ originates from the $\Omega_2$ carrier interaction and can be significantly larger by a factor of $1/\eta$ than the preceding terms. Interestingly, taking into account  the coupling to several excited states, contributions arising from $\ket{A_1}$ and $\ket{A_2}$ (as well as $\ket{E_1}$ and $\ket{E_2}$), both equal $m_s=+1$ and $m_s=-1$ superpositions, have opposite signs and therefore subtract. This limits the maximal amplitude of the uncorrelated flip contributions arising from carrier transitions to $\kappa_1^2\,\Delta$ with $\Delta\sim 4\,GHz$ the excited states energy splitting, that can be considerably lower than expected from the coupling to a single excited state. 

\subsection{Second effective form and gate Hamiltonian}  
In a second stage we consider Hamiltonian\,(\ref{heff1},\,\ref{heff1b}) for two NV centers (k=1,2) and choose $\Delta\epsilon=1/\kappa_2\,\tilde{\Omega}$ with $\kappa_2\ll 1$ denoting the off-resonance of the corresponding transition. That way the phonon transition is only virtually excited which allows to obtain a second effective form  
\numparts
\begin{eqnarray}\label{heff2} H_{\mathrm{eff}}^{\mathrm{II,dp}}&=\sum_{k=1,2} \frac{\tilde{\delta}_k^{\mathrm{dp}}}{2}\,\sigma_x^k-\frac{\Omega_{\mathrm{gate}}}{2} \,\sigma_x^1\,\sigma_x^2\\
\label{heff2b} H_{\mathrm{eff}}^{\mathrm{II,sp}}&=\sum_{k=1,2} \frac{\tilde{\delta}_k^{\mathrm{sp}}}{2}\,\sigma_z^k -\frac{1}{4}\,\frac{\Omega_{\mathrm{gate}}}{2}\,\left(\sigma_x^1\,\sigma_x^2+\sigma_y^1\,\sigma_y^2  \right)
 \end{eqnarray}
\endnumparts
corresponding to the final gate Hamiltonian. Herein the effective gate frequency is defined as $\Omega_{\mathrm{gate}}=\tilde{\Omega}_1\,\tilde{\Omega}_2/\Delta\epsilon$, $\tilde{\delta}_k^{\mathrm{dp}}=\delta_k^{\mathrm{dp}}+(\tilde{\Omega}_1+\tilde{\Omega}_2)\tilde{\Omega}_k/\Delta\epsilon$ and $\tilde{\delta}_k^{\mathrm{sp}}=\delta_k^{\mathrm{sp}}+\frac{(1+2\,\hat{n})\,\tilde{\Omega}_k^2}{4\,\Delta\epsilon}$.\\ For the double-path scheme this corresponds to a $\sigma_x\otimes\sigma_x$-type gate rotating the states within the two-qubit manifolds $\mathcal{M}_1=\{\ket{g_{+1}, g_{-1}}, \ket{g_{-1}, g_{+1}}  \}$ and $\mathcal{M}_2=\{ \ket{g_{+1},  g_{+1}}, \ket{g_{-1},  g_{-1}} \}$ at a Rabi-frequency $\Omega_{\mathrm{gate}} \sim \kappa_1\,\kappa_2\,(\eta\Omega_2)$. Additionally there exist uncorrelated single qubit flips between the ground state levels with $\tilde{\delta}_k^{\mathrm{dp}}\sim \kappa_1\,\eta\Omega_2$ that are by an order $1/\kappa_2$ larger than the gate term itself (see figure\,\ref{b_gatesetup}\,(b)). However both terms commute what allows to remove the single qubit flips by a simultaneous echo-$\pi$-pulse in $\sigma_z$ or $\sigma_y$ on both NV centers, leaving the gate interaction unchanged but adding a negative sign to the uncorrelated single-flip contributions. Interestingly, choosing $\sigma_y$ for that task offers the possibility to decouple the ground state system from decoherence processes as well, therefore extending the coherence time significantly. To conclude, a pure gate interaction can be achieved by adding any periodic pulsed decoupling sequence in $\sigma_y$ acting on an inter-pulse timescale larger than the one required for the effective Hamiltonian form to be valid, i.e. $\Delta t \gg (\Delta\epsilon)^{-1} \sim (\eta\Omega_2)^{-1}$. For a two-qubit $\pi/2$-rotation this gate interaction including the echo-refocusing is illustrated in figure\,\ref{b_gatesim}\,(a). \\
The single-path scheme behaves in a similar way, with the gate interaction  restricted to a rotation in the $\mathcal{M}_1$ manifold, more challenging with respect to the initial state initialization on the nanometer scale of adjacent NV centers that requires individual addressing. For the case of equal $\sigma_z$-contributions, i.e. identical configurations on both of the NV centers, the gate and single qubit contributions commute again and the latter can be removed by an echo pulse and combined with decoupling sequences in $\sigma_x$ and $\sigma_y$, analogue to the two-path situation. Uncorrelated transitions do not occur as they do not form sideband independent paths. Note that dependent on the parameters, the AC Stark shift contributions might be an obstacle in adjusting the off-resonance for the second effective gate Hamiltonian form. Therefore, choosing smaller Rabi frequencies ($\Omega_2\sim \eta\,\kappa\nu$)  might be advantageous to suppress the predominant influence of the carrier term.

\subsection{Comparison to a direct gate implementation}  
At this point it is interesting to compare the relevant timescales of this Raman-induced scheme to a direct gate implementation on the ground-excited state transition. For the direct gate implementation the conditional gate operation is directly implemented between one of the ground and the dissipative excited state, i.e. the Raman transition step is omitted. In the Raman-induced implementation the gate frequency $\propto\kappa_1\,\kappa_2 (\eta\Omega_2)$ and has to be compared to the effective decay rate $\Gamma_{\mathrm{eff}}=\kappa_1^2 \Gamma$ with $\Gamma\simeq 15\,MHz$ the excited state decay rate\,\cite{togan10} suppressed by the probability $\kappa_1^2$ of actually populating the excited level. $\Omega_2$ is limited by the off-resonance to the carrier transition as described above. The corresponding ratio follows as  $\Omega_{\mathrm{gate}}/\Gamma_{\mathrm{eff}} \sim (\kappa_2/\kappa_1)\,(\eta\Omega_2 / \Gamma)$. In contrast to that a direct gate implementation leads to a gate frequency $\Omega_{\mathrm{gate}}^{e\leftrightarrow g}\propto \kappa_1\,\eta\,\Omega$ with the same $\Omega$ limitations, that in this setup has to be related to the bare decay rate $\Gamma$, therefore $\Omega_{\mathrm{gate}}^{e\leftrightarrow g}/\Gamma=\kappa_1\,\eta\Omega/\Gamma$. Note that this is by a factor $\kappa\ll 1$ worse than for the Raman-induced gate scheme.

\subsection{Time-conditioned gate}  
\begin{figure}[htb]
\begin{centering}
\includegraphics[scale=0.45]{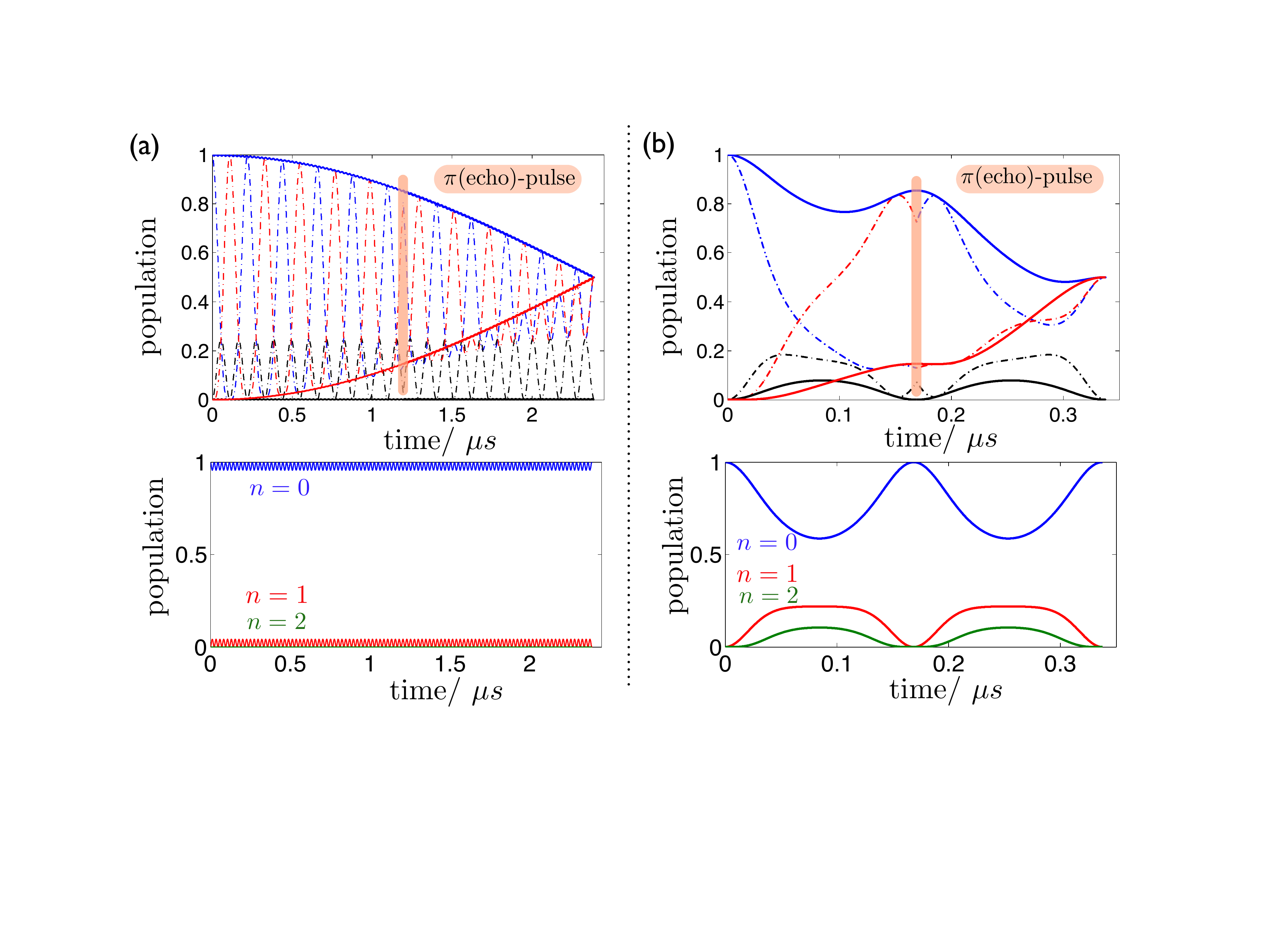}
\caption{\label{b_gatesim} \textbf{Simulated gate interaction} Simulation of the double-path gate interaction\,(\ref{heff2}) for a 15\,nm diamond and \textbf{(a)} $\kappa_1=0.05$, $\kappa_2=0.05$ and $\textbf{(b)}$ $\kappa_1=0.05$, $\kappa_2=1/(2\,\sqrt{2})\simeq 0.35$ (time-conditioned version). The upper figures show the population of the states $\ket{g_{+1},g_{+1}}$ (blue), $\ket{g_{-1},g_{-1}}$ (red) and $\ket{g_{-1}, g_{+1}}, \ket{g_{+1}, g_{-1}}$ (black). Dashed lines illustrate the interaction including the uncorrelated flips that are refocused by an echo pulse, whereas continuous lines focus on the pure gate interaction. The lower plots show the phonon population during the gate interaction that can be significant for the time conditioned (close-resonant) gate.   }
\end{centering}
\end{figure}

To improve this ratio larger $\kappa_2$ values are advantageous, corresponding to smaller off-resonances with respect to the intermediate sideband states. Interestingly the condition $\kappa_2\ll 1$ can be significantly relaxed for the double-path setup, noting that in this case the time evolution following from  Hamiltonians of the form\,(\ref{heff1}) can be exactly integrated\,\cite{roos08} as will be shown in \ref{append_exact_int}. An important prerequisite in that regime consists of compensating the phonon number dependent terms appearing in\,(\ref{heff1}) (the '$\eta^2$-terms') to ensure the commutativity of the gate relevant term (the $\tilde{\Omega}_k$-term) with the contributions that do not involve an effective phonon (de-)\,excitation (the $\delta_k$-contribution). Such a compensation can be achieved by the green compensation couplings illustrated in figure\,\ref{b_gatesetup}\,(a) and also leads to significant improvements for the gate in the perturbative regime\,(\ref{heff2}),\,(\ref{heff2b}) in cases when it is implemented not deep within the $\kappa_2\ll 1$ limit.   Replacing the condition $\Delta\epsilon\ll 1$ by the gate time condition $t_{\mathrm{gate}}=m\,2\pi/\Delta\epsilon$ ($m\in \mathds{N}$), the resulting evolution can still be described by Hamiltonian\,(\ref{heff2}), even this does not hold for intermediate time-steps. That way $\kappa_2=\sqrt{\theta/(2\pi\,m)}$ with $\theta$ the gate rotation angle taking the value $\kappa_2=1/(2\sqrt{2})$ for creating a maximally entangled state ($\theta=\pi/2$) with $m=2$, providing that the phonon population refocuses before the intermediate echo pulse is applied  (see figure\,\ref{b_gatesim}\,(b)).\\
A similar gate that allows one to perform the gate in the non-perturbative regime despite maintaining its independence on the phonon state can be constructed out of the single-path configuration by adding a continuous microwave driving within the ground state triplet manifold\,\cite{bermudez12} and we will discuss that idea in \ref{append_mw_gate}.

\subsection{Size limitation} 
Figure\,\ref{b_gatesetup}\,(c) compares the gate speed to the excited state decay limitation for different configurations including the time conditioned $\kappa_2=0.35$ case. The maximal size limitation is determined by the off-resonance parameter $\kappa_2$ and is given by $\lesssim 25\,nm$ for the $\kappa_2=0.05$ and $\lesssim 35\,nm$ for $\kappa_2=0.1$. Here it should be noted that the lifetime in nanocrystals is increased compared to the bulk counterpart due to the strong change in the refractive index and dielectric screening effects\,\cite{tisler09, greffet11} with typical values\,\cite{beveratos01, neumann09} $\Gamma_{\mathrm{ND}}\simeq 1/2\,\Gamma=7.5\,MHz$. Therefore longer excited state lifetimes have to be expected what was not taken into account in figure\,\ref{b_gatesetup}\,(c) as the exact magnitude is very sensitive to the substrate environment\,\cite{greffet11}. Provided that the Rabi frequencies can take the same values as in the bulk counterpart, a decay rate halved in value would double the ratio $\Omega_{gate}/\Gamma_{eff}$, therefore making the conditional gate operator less prone to spontaneous decay for a given diamond size. For a spherical diamond this would increase the maximal radius by a factor of $\sqrt{2}$.  

\subsection{Influence of dipolar couplings} 
Up to now we neglected the influence of dipolar couplings, that is the optical dipolar coupling\,\cite{lukin00} on the ground-excited state transition as well as the magnetic equivalent\,\cite{albrecht13, bermudez11} within the ground state manifold. Whereas this is a good approximation for specific configurations, e.g. an angle of $54.7^\circ$ of the axis connecting two NV centers to an equally oriented symmetry axis for which the dipolar couplings are exactly zero,  for other configurations they can be of the same magnitude as the gate interaction itself. These couplings take values of $j_{\mathrm{opt}}\simeq 2\pi\cdot52.4\,MHz$ (optical ground excited state coupling) contributing as $\tilde{j}_{\mathrm{opt}}\sim \kappa_1^2\,j_{\mathrm{opt}}$ due to the excited state off-resonance, and $j_{\mathrm{mag}}\simeq 2\pi\cdot 104\,kHz$ (magnetic ground state coupling) for an NV center distance of $r=10\,nm$. That is the effect of the dipolar interactions cannot be neglected in the general case. A detailed discussion about how to include the effect of dipolar couplings in the formalism is presented in \ref{append_dip_coupl}. As a result this requires to replace the detunings by $\epsilon_k\rightarrow \epsilon_k-j_{\mathrm{opt}}/2$  and the gate frequency for the both-path setup by $\Omega_{\mathrm{gate}}^{\mathrm{bp}}\rightarrow \Omega_{\mathrm{gate}}^{\mathrm{bp}}-2\,\tilde{j}_{\mathrm{opt}}-j_{\mathrm{mag}}/2$ and $\Omega_{\mathrm{gate}}^{\mathrm{bp}}\rightarrow \Omega_{\mathrm{gate}}^{\mathrm{bp}}+j_{\mathrm{mag}}/2$ for the $\mathcal{M}_1$-and $\mathcal{M}_2$-interaction, respectively, as well as $\Omega_{\mathrm{gate}}\rightarrow \Omega_{\mathrm{gate}}-4\,\tilde{j}_{\mathrm{opt}}$ for the single-path setup. Herein $\tilde{j}_{\mathrm{opt}}\sim \kappa_1^2 j_{\mathrm{opt}}$. Therefore, as long as $j_{\mathrm{opt}}\ll\nu$ together with identical coupling configurations on both NV centers, both coupling mechanisms can be combined by taking into account the modified detuning configurations and adjusting $\epsilon_k$ correspondingly.  

\subsection{Experimental implementation}
In here we give a brief outline on how we believe the phonon coupling schemes could be realized in experiments. As mentioned earlier the scheme relies on the possibility of individually resolving single excited states, which requires to work in the low temperature regime ($<$10\,K).  A first step naturally consists in the characterization of the mode frequencies, that is, the measurement of an absorption or emission spectrum. As the typical THz phonon frequency range for nanodiamonds is well above the natural linewidth broadening ($\sim$ 15 MHz), we expect a clear significance of phonon sidebands in the spectrum, that allows to determine the relevant frequencies along with a first estimation of the phonon coupling by fitting the data to an appropriate model (see e.g.\,\cite{cirac93}). Beside nanodiamonds on a substrate, recent experiments have demonstrated the levitation of nanodiamonds in optical dipole traps\,\cite{neukirch13}, which is promising in that the phonon mode spectrum can be expected to resemble more closely the one of a free particle as outlined in\,\ref{append_elast_sphere}. Initialization of the electron-spin state in the low temperature regime can be performed by optical pumping using resonant excitation techniques\,\cite{robledo11}. Alternatively, the Lambda scheme, which is also used for the Raman transition in the gate proposals of the previous sections, allows for the initial state preparation in a coherent population trapping configuration\,\cite{togan11}. Readout can be carried out by resonant excitation (projective measurements) on a specific ground-excited state transition\,\cite{robledo11} after mapping the state from the \{$\ket{+1}, \ket{-1} $\} gate-manifold by means of a microwave $\pi$-pulse into \{$\ket{\pm1}, \ket{0}$\}, such as to obtain selective state specific transitions for the read-out process. Here we would like to point out that the individual read-out is challenging on the nanometer scale; however global fluorescence correlation measurements allow for a clear distinction between entangled and mixed states and while full quantum state tomography could be performed analyzing the different fluorescence levels obtained in global measurements\,\cite{dolde13}, the measurement of a small number  of observables sufficies to obtain very tight quantitative bounds on entanglement\,\cite{audenaert06} and fidelities\,\cite{wunderlich09}. Nevertheless, by choosing a configuration of two NV-centers with a distinct symmetry axis orientation, provides, combined with a weak magnetic field, the possibility for individual microwave addressing within the spin-triplet ground state manifold. This also allows, beside the spectrum analysis, for a detailed analysis of the dipolar coupling using double electron-electron resonance techniques (DEER)\,\cite{dolde13}. Here it seems favorable to choose an axes configuration with a dipolar coupling as small as possible in order to make the phonon induced mechanism the dominant one. The gate scheme itself as described in the previous sections relies on purely global laser interactions that do not require individual addressability with the echo $\pi$-pulse implemented by either global microwave or carrier Raman laser couplings, respectively.

\section{Summary}  In summary we analyzed the coupling of nitrogen vacancy centers to long wavelength acoustical phonons, a mechanism capable of mediating gate interactions between NV centers for nanodiamond sizes of several tens of nanometers. Exploiting the existence of a $\Lambda$-scheme in the low temperature and strain limit, fully noise decoupled two qubit  gates can be constructed within the ground state manifold,  even in the presence of dipolar couplings. This might be interesting for the creation of entangled states but also for manipulating the phonon mode itself, that is the control of the motional degrees of freedom, e.g. the cooling of vibrational modes. Moreover, the realization of entangled states in nanodiamonds could have crucial application for future sensing protocols.

\vspace{3ex}

\textit{Author's note: }While finalizing this manuscript we became aware of a similar investigation\,\cite{bennett13} studying phonon induced spin-spin interactions in diamond nanobeams.

\section{Acknowledgements: } We thank M. Aspelmeyer for discussions that contributed to the conception of this project and A. Imamoglu for discussions at early stages of this 
project. This work is supported by the Alexander von Humboldt Foundation,
the BMBF Verbundprojekt QuOReP (FK 01BQ1012), a GIF project and the EU
Integrating Project SIQS and the EU STREP project EQuaM.

\appendix
\section{Calculation of the electron-phonon coupling\,\cite{maze11,doherty11}}\label{append_phon_coupl}
The effect of strain induced by phonons, i.e. the electron-phonon coupling, can be calculated by analyzing the effect of a change in the intra-atom distance on the Coulomb coupling interaction equation\,(\ref{scoulomb1}). Assuming that this displacement is small, a realistic assumption by restricting the analysis to the long wavelength acoustical modes or more precise to the one defined by $k\,l=n\,2\,\pi$ with $n=1$, for which the wavelength is given by the length scale of the diamond crystal, it suffices to expand the coefficients $h_{ij}$ to first order in the atom displacements $\vec{u}_{i}$ out of the equilibrium positions $\vec{r}_i^{\,0}$ (i.e. $\vec{r}_i^{\,0} \rightarrow \vec{r}_i=\vec{r}_i^{\,0}+\vec{u}_i$). With the additional spherical symmetry assumption that the energy will just depend on the absolute value of the relative two atom displacement, that is $h_{ij}=h_{ij}(|\vec{r}_i-\vec{r}_j|)$ one obtains $h_{ij}=h_{ij}^0+\delta h_{ij}$ with
\begin{eqnarray} \delta h_{ij}&=\frac{1}{|\vec{r}_i^{\,0}-\vec{r}_j^{\,0}|}\,\frac{\partial h_{ij}(|\vec{r}_i-\vec{r}_j|)}{\partial\,\, |\vec{r}_i-\vec{r}_j|}\Biggl|_0 \,\left(\vec{r}_i^{\,0}-\vec{r}_j^{\,0}\right)\,\left(\vec{u}_i-\vec{u}_j  \right) \\
&=\frac{1}{|\vec{r}_i^{\,0}-\vec{r}_j^{\,0}|}\,\frac{\partial h_{ij}(|\vec{r}_i-\vec{r}_j|)}{\partial\,\, |\vec{r}_i-\vec{r}_j|}\Biggl|_0 \,\left(\vec{r}_i^{\,0}-\vec{r}_j^{\,0}\right)\,\mathbf{e}\left(\vec{r}_i^{\,0}-\vec{r}_j^{\,0}\right)
\end{eqnarray}
where in the second line we introduced the displacement tensor $\mathbf{e}_{\mu\nu}=\partial u_\mu / \partial r_\nu$ and the derivation is evaluated at the equilibrium position of the atoms. Inserting the explicit expressions for the positions $\vec{r}_i^{\,0}$ as defined in figure\,\ref{b_levelstrain}\,(a) and using the expressions of the electronic states $\mathcal{M}_{\mathrm{el}}$ in terms of the dangling bonds orbitals $\mathcal{M}_{\mathrm{db}}$,  the strain Hamiltonian\,(\ref{scoulomb2}) can be rewritten as
\begin{eqnarray}\label{appendV1} \eqalign{\delta V_{\mathcal{M}_{\mathrm{el}}}=-2\,&\zeta\,\mathbf{e}_{xx} \,\ket{e_x}\bra{e_x}-2\,\zeta\,\mathbf{e}_{yy}\,\ket{e_y}\bra{e_y} -8\,\beta^2\,\zeta\,\mathbf{e}_{zz}\,\ket{a_2}\bra{a_2}\\&-\zeta\,(\mathbf{e}_{xy}+\mathbf{e}_{yx})\,\left(\ket{e_x}\bra{e_y}+\mathrm{h.c.} \right)\,. } \end{eqnarray}
Herein we defined 
\begin{equation} \zeta=\sqrt{\frac{2}{3}}\,\frac{\partial h_{ij}(|\vec{r}_i-\vec{r}_j|)}{\partial\,\, |\vec{r}_i-\vec{r}_j|}\Biggl|_{0,C}\, q      \end{equation}
with the index $C$ referring to the coupling for two carbon atoms and the difference for the carbon-nitrogen case is accounted for by the factor $\beta$. The quantity $q$ denotes the next neighbour distance in the diamond lattice and is equal to $q=\sqrt{3/8}\,(\vec{r}_i^{\,0}-\vec{r}_j^{\,0})$. Moreover couplings between $\ket{a_2}$ and $\ket{e_{x,y}}$ levels have been neglected, justified by the large energy separation. Those would correspond to strain induced transitions between the ground and excited states of the NV center and consequently do not play a significant role. Note also that the energy level $\ket{a_1}$ has been neglected as it is delocalized in the valence band and does not contribute to the properties of the NV center energy level structure. \par
Out of equation\,(\ref{appendV1}) the impact of strain on the NV center energy levels can be calculated by using the expression of the energy levels in terms of the electronic states $\mathcal{M}_{\mathrm{el}}$ as provided e.g. in\,\cite{maze11}. For the `two hole' description the strain perturbation Hamiltonian takes the form 
\begin{equation}\label{app_elph1} H_{\mathrm{el-phon}}= \left[  \delta V_{\mathcal{M}_{\mathrm{el}}}\otimes \mathds{1}_{\mathrm{el}}+\mathds{1}_{\mathrm{el}}\otimes \delta V_{\mathcal{M}_{\mathrm{el}}}  \right]\otimes \mathds{1}_{\mathrm{spin}}   \end{equation}
with $\mathds{1}_{\mathrm{el}}$ the identity on $\mathcal{M}_{\mathrm{el}}$ and $\mathds{1}_{\mathrm{spin}}$ the one on the spin degrees of freedom. Projecting Hamiltonian\,(\ref{app_elph1}) on the NV center energy level states finally leads to
\begin{equation}\label{a_gs} H_{\mathrm{el-phon}}^{\mathrm{gs}}=2\,\delta_1\,\left( \ket{g_0}\bra{g_0}+\ket{g_{\mathrm{+1}}}\bra{g_{\mathrm{+1}}}+\ket{g_{\mathrm{-1}}}\bra{g_{\mathrm{-1}}}  \right)   \end{equation}
for the ground state electron spin triplet states and
\begin{equation}\fl \label{a_es} H_{\mathrm{el-phon}}^{\mathrm{es}}=\left(\begin{array}{*{20}{c}} \delta_1+\delta_4 &0 &0 &0 &\delta_2 & -i\,\delta_3   \\
0 & \delta_1+\delta_4  & 0 & 0 & i\,\delta_3 & -\delta_2\\
0 & 0  & \delta_1+\delta_4+\delta_2 & \delta_3 & 0 & 0\\
0 & 0  & \delta_3 & \delta_1+\delta_4-\delta_2 &0 & 0\\
\delta_2 & -i\,\delta_3  & 0 & 0 &\delta_1+\delta_4 &0 \\
i\,\delta_3 & -\delta_2  & 0 & 0 &0 &\delta_1+\delta_4
\end{array}\right)    \end{equation}
for the excited state triplet states in the basis $\{ \ket{A_1},\ket{A_2},\ket{E_x}, \ket{E_y}, \ket{E_1}, \ket{E_2} \}$ with $\delta_1=-\zeta\,(e_{xx}+e_{yy})$, $\delta_2=-\zeta\,(e_{xx}-e_{yy})$, $\delta_3=-\zeta\,(e_{xy}+e_{yx})$ and $\delta_4=-8\beta^2\,\zeta\,e_{zz}$. Combining the Hamiltonians (\ref{a_gs}) and (\ref{a_es}) and neglecting the off-resonant couplings between $\{ \ket{A_1}, \ket{A_2} \}\leftrightarrow \{ \ket{E_1}, \ket{E_2} \}$ results in  Hamiltonian\,(\ref{dhstrain}) of the main text. \par
In the case of phonons the displacement can be expressed in terms of Bloch type wavefunctions, such that for a specific mode $\alpha$, and noting that in the framework of classical elasticity theory applicable in the long wavelength limit the microscopic structure can be replaced by a continuous displacement field ($\vec{r}_i \rightarrow \vec{r}$), the displacement takes the form
\begin{equation} \vec{u}_\alpha(\vec{r}) =\sqrt{\frac{\hbar}{2\,M\,\nu(\vec{k})}}\,\vec{e}^{\,(\alpha)}\,\left[  a_\alpha e^{i\,\vec{k}\vec{r}}+ a_\alpha^\dagger e^{-i\,\vec{k}\vec{r}}  \right]   \end{equation}
with $M$ the total mass of the system, $\vec{e}^{\,(\alpha)}$ the eigenvector of mode $\alpha$, $\vec{k}$ the wavevector and $\nu(k)$ the angular frequency and $a_\alpha$, $a_\alpha^\dagger$ the mode annihilator and creator operators, respectively. With the definition of $\mathbf{e}_{\mu\nu}=\partial u_\mu / \partial r_\nu$ and using that $\vec{k}\,\vec{r}\ll 1$ this results in equation\,(\ref{modedispl}) in the main text, that together with the electron-phonon coupling Hamiltonians\,(\ref{a_gs}) and $(\ref{a_es})$ complete the analysis of the phonon influence on the NV energy level structure.

\section{Phonon-Coupling in the Elastic Sphere Model}\label{append_elast_sphere}
In here we will reconsider the NV-phonon coupling (deformation potential coupling) for the special case of a sphere subject to stress-free boundary conditions. Confinement effects in such finite systems lead to a modification of the phonon modes compared to the periodic boundary condition analysis provided earlier in section\,\ref{sect_phon_coupl}. This modification depends on the shape and size, or more precise on the boundary conditions of the particle under consideration.

 The acoustical vibrations $\mathbf{u}$ of a homogeneous, free spherical elastic body in the framework of continuous elasticity theory can be described by
\begin{equation}\label{lamb1}  \frac{\partial^2}{\partial t^2}\,\mathbf{u}(\mathbf{r},t)=\frac{\lambda+\mu}{\rho}\,\nabla\,\left(\nabla\cdot\mathbf{u}(\mathbf{r},t)\right)+\frac{\mu}{\rho}\,\nabla^2\mathbf{u}(\mathbf{r},t)  \end{equation}
and has been first studied by Lamb\cite{lamb1882}. Herein $\lambda$ and $\mu$ are the Lame's constants that describe the material-dependent elastic properties and are related to the transverse and longitudinal speed of sound in diamond by $v_t=\sqrt{\mu/\rho}=1.283\cdot 10^4\,{\rm(m/s)}$ and $v_l=\sqrt{\lambda+2\mu/\rho}=1.831\cdot 10^4\,{\rm(m/s)}$, respectively, with $\rho=3.512\,g/cm^3$  the mass density\,\cite{mcskimin72}.  This continuous elastic body model has been successfully applied to describe the phonon properties of nanoparticles as validated by numerous experiments, and forms a good description as long a the particle size is not too small, that is, as long as the phonon wavelength is much larger than the interatomic distance to allow for the homogeneous continuum description. 

\begin{figure}[htb]
\begin{centering}
\includegraphics[scale=0.4]{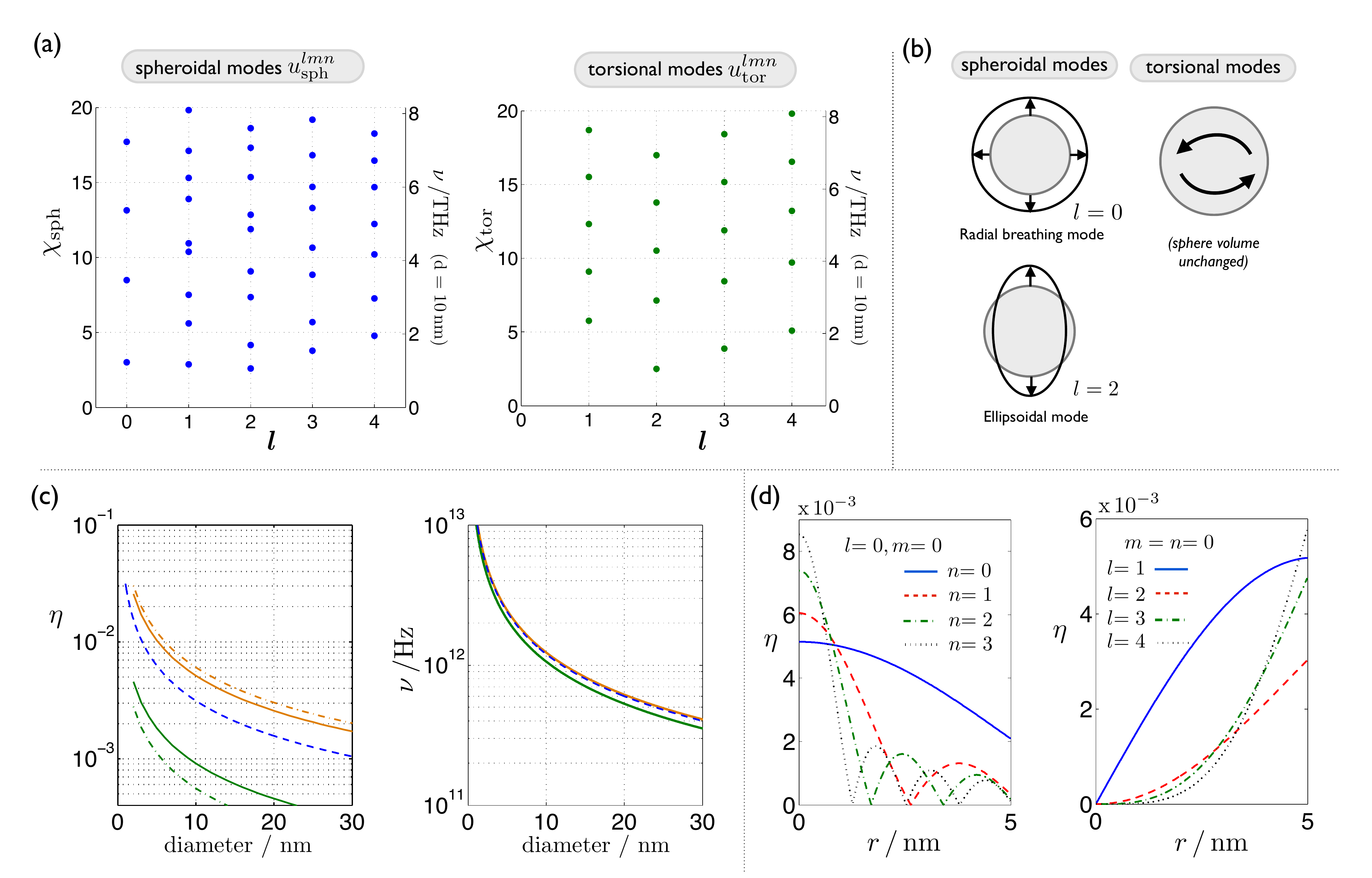}
\caption{\label{b_lamb} \textbf{Mode properties and NV-phonon coupling in the elastic sphere model: } \textbf{(a)} Dimensionless eigenvalues $\chi=kR$ of the spheroidal and torsional modes in diamond. The modes are characterized by the orbital quantum number $l$ and exhibit a (2$l$+1)-degeneracy in $m$. Higher modes for a given $l$ account for the overtones denoted by $n$. For a diamond of 10\,nm in diameter the frequencies following out of\,(\ref{lamb6}) are indicated by the right axis. \textbf{(b)} Schematic illustration of spheroidal (the breathing mode and the lowest energy mode $l$=2) and torsional modes. \textbf{(c)} \textit{Left: } Scaling of the phonon coupling coefficient $\eta$ with the particle diameter. The blue dashed line corresponds to the expectation for the lowest frequency mode with periodic boundary conditions and corresponds to the one of figure\,\ref{b_levelstrain}, whereas orange and green lines follow out of the elastic sphere model for $l$=$m$=0,\,$n$=0 (orange, solid line), $l$=$m$=0,\,$n$=1 (orange, dashed dotted) both at r=0, and $l$=2,\,$m$=0,\,$n$=0 (green, solid line), $l$=2, $m$=$\pm$1,\,$n$=0 (green, dashed dotted) both at r=R/2 evaluated at the angles of maximal coupling, respectively. All lines exhibit the expected $\eta\propto 1/R$ behaviour. \textit{Right: } Mode frequency scaling with diameter for the lowest mode out of periodic boundary conditions (blue, dashed), the elastic sphere breathing mode ($l$=$m$=$n$=0; orange) and the lowest energy elastic sphere ellipsoidal mode ($l$=2, $m$=$n$=0; green). \textbf{(d)} Radial coupling distributions for a spherical diamond of $R=5\,nm$ and different modes as indicated in the figure. Whereas the $l$=0 modes are completely angle independent, the $l\neq0$, $m$=0 modes are evaluated at the angle $\theta=0$ of maximal coupling (independent of $\phi$).     }
\end{centering}
\end{figure}

The equation of motion\,(\ref{lamb1}) can be solved by introducing a scalar $\phi\sim\psi_{lm}(hr,\Omega)$ and vector potential $\mathbf{A}\sim\mathbf{r}\,\psi_{lm}(kr,\Omega)=\hat{\mathbf{e}}_r\,r\,\psi_{lm}(kr,\Omega)$ with $\psi_{lm}(kr,\Omega)=j_l(kr)\,Y_{lm}(\Omega)$, wherein $j_l(kr)$ and $Y_{lm}(\Omega)$ are the $l$-th order spherical Bessel function and the (l,m)-spherical harmonics, respectively\,\cite{eringen75, tamura82}. The corresponding displacement modes  follow as derivatives of those potentials, with the scalar potential describing compressive (longitudinal) and the vector potential shear (transverse) waves, and can be classified into torsional $\mathbf{u}_{\rm tor}^{l,m,n}$ and spheroidal modes $\mathbf{u}_{\rm sph}^{l,m,n}$\,(see figure\,\ref{b_lamb}\,(b)). Herein the  orbital quantum number $l$ and its z-component $m$ with $|m|\leq l$ characterize the mode's spherical symmetry whereas $n$ refers to the specific overtone. Imposing stress-free boundary conditions results in an eigenvalue equation that allows to determine the $k-$ and $h$-values along with the mode frequencies.\par
The \textit{torsional modes} are characterized by a purely transversal character without radial displacement, that is the sphere volume remains unchanged under these vibrations
\begin{equation} \mathbf{u}_{\rm tor}^{l,m,n}=\mathcal{N}_{l,m,n}^{\rm tor}\,\nabla\times\left[\hat{\mathbf{e}}_r\,r\,\psi_{l,m}(kr,\Omega)\right]\quad, \qquad l\geq 1  \end{equation}
with $\mathcal{N}_{l,m,n}^{\rm tor}$ describing a normalization constant. The eigenvalue equation in that case takes the form
\begin{equation} (l-1)\,j_l(\chi)-\chi\,j_{l+1}(\chi)=0  \end{equation}
with the definition $\chi=k\,R$ and $R$ being the sphere radius. Note that this equation is independent of the particle's elastic constants and is therefore of universal character.\par
In contrast to that, \textit{spheroidal modes} exhibit a mixed longitudinal and transversal character and are described by
\begin{equation}\fl \mathbf{u}_{\rm sph}^{l,m,n}=\mathcal{N}_{l,m,n}^{\rm sph}\,\left[ p_{l}\,\left(\frac{1}{k}\,\nabla \psi_{lm}(hr,\Omega)   \right)+q_l\,\left(\frac{1}{k}\,\nabla\times\nabla\times\left[\hat{\mathbf e}_r\,r\,\psi_{l m}(kr,\Omega)\right]\right)  \right]   \end{equation}
with the coefficients $p_l$ and $q_l$ following out of ($\chi=k\,R$, $\xi=h\,R$)
\begin{eqnarray}\eqalign{ \left(\begin{array}{*{20}{c}}  \alpha_l & \beta_l \\ \gamma_l & \delta_l  \end{array}\right) \,\left(\begin{array}{*{20}{c}} p_l \\ q_l  \end{array}\right)=0 \\ \fl  {\rm with}\quad \begin{array}{*{20}{c}}  \alpha_l=-(\chi^2/\xi) j_l(\xi)+2(l+2)\,j_{l+1}(\xi)\,,\\\gamma_l= -(\chi^2/\xi)\,j_l(\xi)+2(l-1)j_{l-1}(\xi)\,, \end{array} \quad \begin{array}{*{20}{c}}  \beta_l=l\chi\,j_l(\chi)-2l(l+2)j_{l+1}(\chi)\,, \phantom{dsekr}\\ \delta_l=(l+1)\left[ 2(l-1)j_{l-1}(\chi)-\chi j_l(\chi) \right]\,, \end{array}} \end{eqnarray}
and $q_l=0$ for $l=0$ (pure radial displacement, 'breathing mode'). The coefficients $\chi$ and $\xi$ are obtained from the eigenvalue equations $\alpha_l\delta_l-\beta_l\gamma_l=0$ $(l\neq0)$ and $\alpha_l=0$ $(l=0)$ by introducing the material dependent relation $\xi=(v_t/v_l)\,\chi$.\par
The mode eigenvalues $\chi$ calculated that way for diamond are depicted in figure\,\ref{b_lamb}\,(a) together with the corresponding frequency $\nu$ for a diamond of 10\,nm in diameter, the latter following from the relation
\begin{equation}\label{lamb6}   \nu=v_t\,k=v_l\,h=v_t\,\frac{\chi}{R}\,. \end{equation}
Note that these modes are degenerate in $m$ for a perfect spherical symmetry. Out of\,(\ref{lamb6}) it follows that the frequency scales as the inverse of the radius, that is it exhibits the same scaling as the one obtained from the periodic boundary condition calculation. Moreover we show in figure\,\ref{b_lamb}\,(c) that the magnitude for the lowest breathing mode (l=0) is in good agreement with the lowest mode obtained in the periodic approach, that will turn out to form a promising mode for the NV-center coupling. 

We will now proceed by calculating the coupling (deformation potential coupling) to these elastic sphere modes. In a second quantized form the displacement takes the form\,\cite{takagahara96}
\begin{equation}\label{lamb7} \mathbf{u}(r,\Omega)=\sum_{lmn,\tau}\sqrt{\frac{\hbar}{2\rho\nu_{lmn}^\tau}}\,(a_{l,m,n}^\tau+(-1)^m {a_{l,-m,n}^{\tau\dagger}})\,\mathbf{u}_{\tau}^{l,m,n}  \end{equation} 
with $\tau$ denoting either spheroidal or torsional modes and the displacement is normalized over the crystal volume as $\int\mathrm{d}^3 r\,(\mathbf{u}_{\tau}^{l,m,n})^*\cdot\mathbf{u}_{\tau}^{l,m,n} =1 $. Combining\,(\ref{lamb7}) with\,(\ref{dhstrain}) allows for a straightforward calculation of the NV-center coupling to a specific mode. For the excited states $\ket{e}\in \{\ket{A_1}, \ket{A_2}, \ket{E_1}, \ket{E_2} \}$ the phonon coupling\,(\ref{dhstrain}) can be approximated by the rather simple expression
\begin{equation}\label{lamb8} H_{\rm{el-phon}}^\tau\simeq \zeta\,\left( e_{xx}+e_{yy}+e_{zz} \right)\,\ket{e}\bra{e}=\zeta\,{\rm div}(\mathbf{u}_\tau^{l,m,n})\ket{e}\bra{e} \end{equation}
such that for the spheroidal modes
 \begin{equation}\fl H_{\rm{el-phon}}^{\rm sph} \simeq -\zeta\,\sqrt{\frac{\hbar}{2\rho\nu_{lmn}^{\rm sph}}}\mathcal{N}_{l,m,n}^{\rm sph} p_l\,h_{l,n} j_l(h_{l,n}\,r) Y_{l,m}(\Omega) \left(a_{l,m,n}^{\rm sph}+(-1)^m a_{l,-m,n}^{{\rm sph}\dagger}\right)\ket{e}\bra{e} \end{equation}
and $H_{\rm{el-phon}}^{\rm tor}\simeq 0$ for the torsional modes, with $h_{l,n}=(v_t/v_l)\,(\chi_{l,n}/R)$. This is in accordance with the dominant contribution expected for the general case of a deformation potential coupling\,\cite{takagahara96}. From\,(\ref{lamb8}) it follows that only the longitudinal displacement contributes significantly to the NV-center phonon-coupling mechanism and therefore the coupling to torsional modes can be neglected. It should be noted that $H_{\rm el-phon}\propto \mathcal{N}_{l,m,n}^{\rm sph}\,h_{l,n}/\sqrt{\omega_{lmn}}\sim R^{-2} $ scales quadratically with the inverse particle radius and therefore the coupling factor $\eta$ as defined via\,(5) scales as $\eta\sim R^{-1}$, thus exhibiting the same scaling as already obtained in section\,\ref{sect_phon_coupl}. We verified that scaling behavior in figure\,\ref{b_lamb}\,(c) for the breathing mode (l=m=0) and the lowest frequency mode (l=2,m=0,$\pm$1). The breathing mode coupling constant $\eta$ matches fairly well the expectation calculated by using periodic boundary conditions. Here one should take into account as well the radial position dependence that is illustrated in more detail in figure\,\ref{b_lamb}\,(d): Only the $l=0$ modes have a non-vanishing coupling around the particle center (r=0) whereas modes with $l\neq 0$ exhibit an increasing region of vanishing coupling for increasing $l$ around the particle center (for a fixed overtone number $n$). Combined with the fact that NV-center near the surface are less stable and additionally are more prone to decoherence, the $l=0$ breathing modes can be considered as the promising modes to obtain phonon-coupling, in particular the low frequency $n=0$ mode that shows the most uniform coupling achievable throughout the possible NV-center positions within the crystal. However one should keep in mind that the coupling to other modes is, dependent on the specific position of the NV center within the diamond, not necessarily weaker than the coupling to $l=0$ and in figure\,\ref{b_lamb}\,(c) this behaviour just arises from the fact that for the $l=0$ mode the NV-center is assumed to be at the center (the position of maximal coupling) whereas for the $l=2$ modes the reasonable assumption $r=R/2$ has been chosen, that does not correspond to the maximal coupling position which in fact would be given by $r=R$. As a general behaviour higher overtones, also known as \textit{inner modes}, are accompanied with a decreasing phonon coupling layer around the surface (as can e.g. be seen in figure\,\ref{b_lamb}\,(d)\textit{left} and a similar behaviour would be observed for $l\neq0$ modes); these modes depend only weakly on the specific boundary conditions.\par
In summary, the general scaling and magnitude of the phonon coupling obtained previously in section\,\ref{sect_phon_coupl} by assuming periodic boundary conditions (`infinite crystal approximation') is in good agreement with the results obtained by assuming a confined finite spherical particle. This has indeed to be expected as the scaling properties arise from universal dimensionality arguments as e.g. the mode normalization factor, thus allowing a rather simple estimation of the coupling properties and strength even for different `shapes' (dimensionality) in the periodic boundary model. However the exact microscopic spatial coupling, e.g. the spatial distribution of the coupling parameter,  depends on the explicit mode properties for which the particle confinement and shape become crucial and in fact the low frequency modes (n=0 modes) are most sensitive to a change in these surface properties.

\section{Exact integration of the gate Hamiltonian}\label{append_exact_int}
The time evolution following out of Hamiltonian\,(\ref{heff1}) can be integrated exactly for commuting state operators (e.g. for the double-path gate setup), what allows to overcome the $\kappa_2\ll 1$ limit by identifying appropriate time conditions. We will consider the generic form
\begin{equation}\label{shex1} H(t)=i\,(\gamma(t)\,\hat{O}\,a^\dagger-\gamma^*(t)\,\hat{O}^\dagger\,a)  \end{equation} 
with $[ \hat{O}, \hat{O}^\dagger]=0$.
For the double-path setup considered here\,(\ref{heff1})
\begin{equation}  \gamma(t)=\frac{\tilde{\Omega}}{2}\,e^{i\,\Delta\epsilon\,t} \end{equation}
and
\begin{equation}\label{exint3} \hat{O}=\sigma_x^1+\sigma_x^2+2\,\mathds{1}\,. \end{equation}
Using the properties of the displacement operator $D(\alpha)=\exp(\alpha\,a^\dagger-\alpha^* a)$, the time evolution of\,(\ref{shex1}) follows as\,\cite{roos08}
\begin{equation} U=\mathcal{T}\,e^{-i\,\int H(t')\,\mathrm{d}t'}=D(\alpha(t)\,\hat{O})\,\exp\left( \frac{1}{2} \,\left[\beta(t)\,\hat{O}\,\hat{O}^\dagger-\beta^*(t)\,\hat{O}^\dagger\,\hat{O}  \right] \right)   \end{equation}
with
\begin{eqnarray} \alpha(t)&=\int_0^t\,\mathrm{d}t'\,\gamma(t')=-\frac{i}{2}\,\frac{\tilde{\Omega}}{\Delta\epsilon}\,\left( e^{i\,\Delta\epsilon\,t}-1 \right) \\
\beta(t)&=\int_0^t\mathrm{d}t'\,\gamma(t')\int_0^{t'}\mathrm{d}t''\,\gamma^*(t'')=i\,\frac{\tilde{\Omega}^2}{4\Delta\epsilon}\,\left( t+\frac{i}{\Delta\epsilon}\,\left[  e^{i\,\Delta\epsilon\,t}-1\right] \right)\,.
 \end{eqnarray}
Noting that the phonon dependence only appears in the displacement operator, it can be eliminated by choosing the gate time $t_{\mathrm{gate}}$ such that $\Delta\epsilon\,t_{\mathrm{gate}}=m\cdot 2\pi$ with $m\in \mathds{Z}$ in which case $\alpha(t_{\mathrm{gate}})=0$. Implying that condition is fulfilled, the total evolution corresponds exactly to the one out of\,(\ref{heff2}) (up to local contributions of the first order effective Hamiltonian and global phases), i.e.
\begin{equation}
U(t_\mathrm{gate})=\exp\left(-i\,\left[ -\frac{\tilde{\Omega}^2}{2\,\Delta\epsilon}\,\left(\sigma_x^1\,\sigma_x^2+2\,\left[  \sigma_x^1+\sigma_x^2\right]\right) \right]\,t_{\mathrm{gate}}  \right) \, . \end{equation}

\section{Microwave assisted gate}\label{append_mw_gate}
\begin{figure}[htb]
\begin{centering}
\includegraphics[scale=0.4]{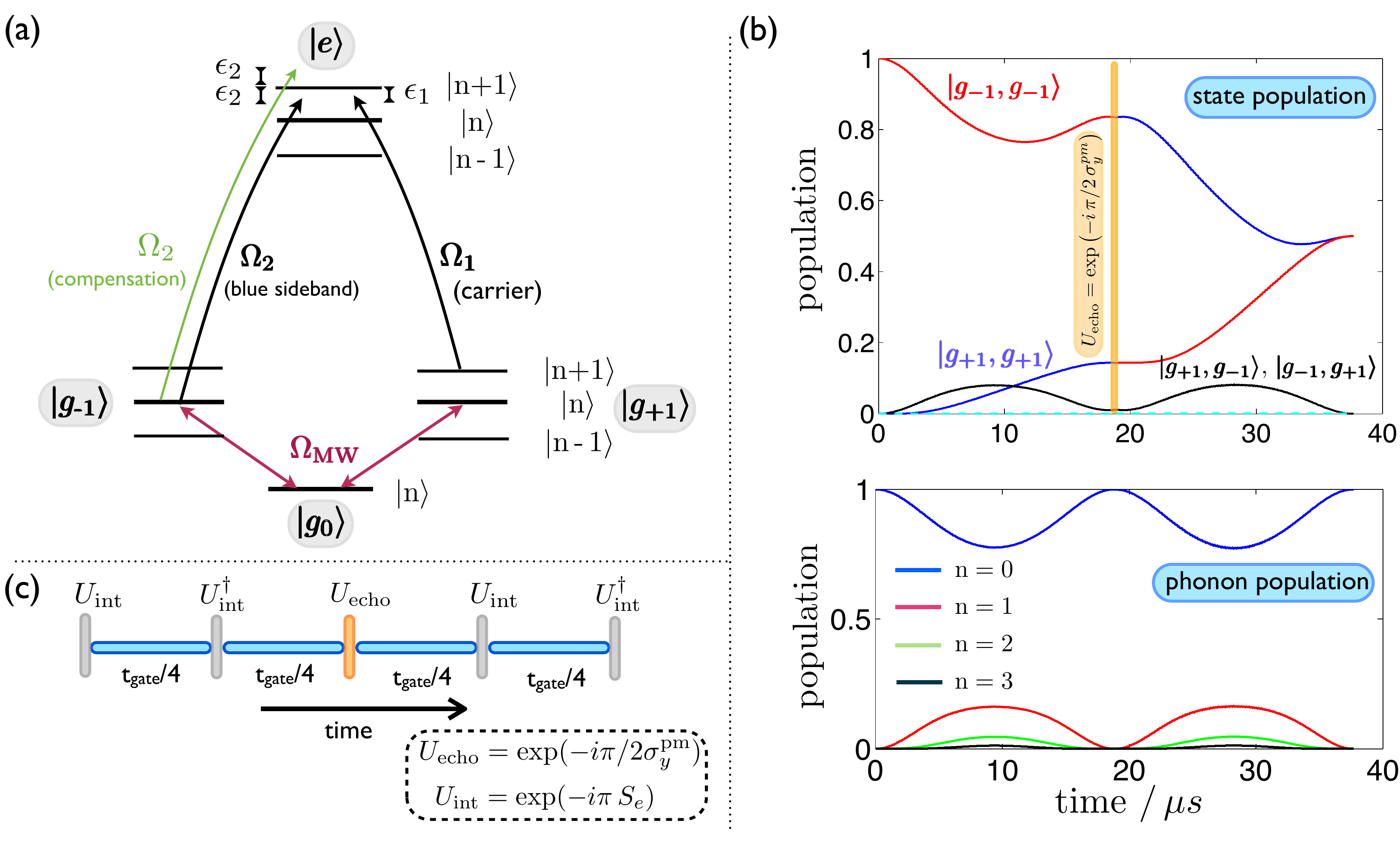}
\caption{\label{b_mw_gate} \textbf{Microwave assisted gate: } \textbf{(a)} Gate setup. The gate consists of a single blue sideband transition connecting $|g_{\mathrm{-1}}\rangle$ to $\ket{g_{\mathrm{+1}}}$ (realized by a Raman transition via the excited state) and a continuous microwave coupling $\Omega_{\mathrm{MW}}$ between the states $|g_{0}\rangle$ and $\ket{g_{\pm1}}$. An additional compensation coupling (green line) removes the phonon number dependent ac-Stark-shift. \textbf{(b)} Simulated gate process for $\Omega_{MW}=10\,MHz$, $\Omega_1=\eta\Omega_2=6.3\,MHz$, $\epsilon_1=0.4\,GHz$, $\Delta\epsilon=53\,MHz$ and a diamond size of $15\,nm$. The time condition is given by $t_{\mathrm{gate}}=4\pi/\Delta\epsilon$ (n=2). For simplicity and clarity the carrier contributions of the blue sideband laser have been neglected in the simulation. The cyan dashed line corresponds to the sum of all populations including the state $|g_{0}\rangle$. \textbf{(c)} Possible implementation of the gate process. Blue lines denote a time evolution under the Hamiltonian form\,(\ref{hmw1}) and $S_e\in\{\sigma_x^{\mathrm{pm}}, S_y, S_z \}$.  }
\end{centering}
\end{figure}

A gate interaction with similar properties as the double-path gate discussed in the main text can be constructed out of the single-path setup combined with a continuous microwave driving of the states $\ket{g_0}\leftrightarrow \ket{g_{+1}}$ and $\ket{g_0}\leftrightarrow \ket{g_{-1}}$ (see figure\,\ref{b_mw_gate}\,(a)). Mainly this allows to perform the gate in the non-perturbative regime implying a time condition to assure the independence of the phonon state. As a side effect such a driving also decouples the gate from ground state decoherence. \par
Considering the single path setup alone, the gate term in\,(\ref{heff1b}) includes the operators
\begin{equation}   \sigma_{\pm}=\frac{1}{2}\,\left( \sigma_x\pm i\,\sigma_y \right)\,. \end{equation}
 In that case a closed phase space trajectory, that is a refocusing to the initial phonon state  at a specific time, is prevented by the rotation around two orthogonal axes ($\sigma_x$ and $\sigma_y$)\,\cite{bermudez12}. In a more formal way the non-commutativity of $\sigma_x$ and $\sigma_y$ prevents the exact integration of the gate Hamiltonian as described in \ref{append_exact_int}. Now adding a continuous microwave driving such that the $\sigma_y$ contribution is suppressed, removes those difficulties, and additionally leads to a $(\sigma_x\otimes\sigma_x)$-type gate similar to the double-path gate proposal that rotates states both within $\mathcal{M}_1$ and $\mathcal{M}_2$. One possibility to achieve that task would be to continuously drive the states $\ket{g_{+1}}\leftrightarrow \ket{g_{-1}}$\,\cite{bermudez12, lemmer13}, what however is not really practicable for our setup. Therefore we will incorporate the full ground state triplet and show that a driving of the form $\ket{g_{0}}\leftrightarrow \ket{g_{\pm 1}}$ will be suitable for this task as well. In the following we will refer to the qubit operators within $\{ \ket{g_{+1}}, \ket{g_{-1}} \}$ by the indices `pm' and denote the Pauli spin-1/2 operators in that manifold as $\{\sigma_x^{\mathrm{pm}},\sigma_y^{\mathrm{pm}}, \sigma_z^{\mathrm{pm}}, \mathds{1}_{\mathrm{pm}}\}$ whereas we will denote the spin-1 operators in the ground-state triplet manifold as $\{ S_x, S_y, S_z, \mathds{1} \}$.\par
The Hamiltonian of the total system including the microwave driving $H_{MW}$ is given by
\begin{equation}\label{hmw1} H=\sum_{j=1,2}H_{MW}^{(\mathrm{j})}+H_{\mathds{1}_{\mathrm{pm}}}^{(\mathrm{j})}+H_{\sigma_z^{pm}}^{(\mathrm{j})}+H_{\mathrm{gate}}^{(\mathrm{j})}  \end{equation}
with
\begin{eqnarray}\label{mw_ham}\eqalign{\fl
H_{MW}^{(\mathrm{j})}&=\frac{\Omega_{MW}}{2}\,S_x^{(\mathrm{j})}    \\\fl
H_{\mathds{1}_{\mathrm{pm}}}^{(\mathrm{j})}&=\frac{\delta_{\mathds{1}}}{2}\,\mathds{1}_{\mathrm{pm}}^{(\mathrm{j})} \qquad \qquad\qquad\qquad\quad\,\, \mathrm{with} \quad \delta_{\mathds{1}}=\frac{1}{4}\,\left(\frac{\Omega_1^2}{\epsilon_1}+\frac{\Omega_2^2}{\nu+\epsilon_2}\right)\\\fl
H_{\sigma_z^{pm}}^{(\mathrm{j})}&=\frac{\delta_{\mathrm{sp}}}{2}\,\sigma_z^{\mathrm{pm},(\mathrm{j})} \qquad\qquad\qquad\qquad \mathrm{with}\quad \delta_{\mathrm{sp}}=\frac{1}{4}\,\left( \frac{\Omega_1^2}{\epsilon_1}-\frac{\Omega_2^2}{\nu+\epsilon_2} \right) \\\fl
H_{\mathrm{gate}}^{(\mathrm{j})}&=i\,a^\dagger\,\frac{\tilde{\Omega}}{2}\,e^{i\,\Delta\epsilon\,t}\,\frac{1}{2}\,\left(\sigma_x^{\mathrm{pm}}+i\sigma_y^{\mathrm{pm}}\right)+\mathrm{h.c.} \quad \mathrm{with}\quad \tilde{\Omega}=\frac{1}{4}\,\Omega_1\,(\eta\Omega_2)\,\left(\frac{1}{\epsilon_1}+\frac{1}{\epsilon_2}  \right)}
  \end{eqnarray}

where we assumed that the $\eta^2$-terms have been successfully compensated to avoid the existence of phonon-number dependent terms (see green laser coupling in figure\,\ref{b_mw_gate}\,(a)). This describes the microwave driving, the global ac-Stark shift of the $\ket{g_{\pm 1}}$ states with respect to the $\ket{g_0}$ state, the relative shift of the $\ket{g_{\pm 1}}$ states and the gate relevant term, respectively. Note that the global $H_{\mathds{1}_{\mathrm{pm}}}$ term can be neglected if the microwave frequency is already tuned to the ac-Stark shifted states.\par
In an interaction picture with respect to the continuous microwave driving and assuming that $\Omega_{MW}\gg \{\tilde{\Omega}, \delta_{sp},\delta_{\mathds{1}}  \}$, the following substitutions can be made by neglecting fast rotating terms (`rotating-wave approximation')
\begin{eqnarray} \label{mw_subst} 
\eqalign{\fl\sigma_x=S_x S_x- S_y S_y \quad  & \rightarrow \quad S_x S_x -\frac{1}{2}\,\left(S_y S_y+S_z S_z  \right)=\frac{3}{4}\,\sigma_x^{\mathrm{pm}}-\frac{1}{4}\,\mathds{1}_{\mathrm{pm}}+\frac{1}{2}\,\ket{g_{0}}\bra{g_{0}}\\ 
\fl\sigma_y=S_x S_y+S_y S_x  \quad &\rightarrow \quad  0\\
\fl\mathds{1}_{pm}=S_z^2\quad  &\rightarrow \quad \frac{1}{2}\,\left(S_z S_z+ S_y S_y \right)=-\frac{1}{4}\,\sigma_x^{\mathrm{pm}}+\frac{3}{4}\,\mathds{1}_{\mathrm{pm}}+\frac{1}{2}\ket{g_{0}}\bra{g_0}\\
\fl\sigma_z^{pm}=S_z  \quad &\rightarrow \quad 0\,.}
\end{eqnarray}
Therefore, in that frame, the Hamiltonian takes the form
\begin{eqnarray}\label{hmwint} \eqalign{ H_{\mathrm{int}}\simeq&\sum_{i=1,2}\frac{\delta_{\mathds{1}}}{2}\left(-\frac{1}{4}\,\sigma_x^{\mathrm{pm}}+\frac{3}{4}\,\mathds{1}_{\mathrm{pm}}+\frac{1}{2}\ket{g_{0}}\bra{g_0} \right)_{(i)} \\
&+\left[i\,a^\dagger\,\frac{\tilde{\Omega}}{2}\,e^{i\,\Delta\epsilon\,t}\,\frac{1}{2}\left(  \frac{3}{4}\,\sigma_x^{\mathrm{pm}}-\frac{1}{4}\,\mathds{1}_{pm}+\frac{1}{2}\,\ket{g_{0}}\bra{g_{0}}\right)_{(i)}+\mathrm{h.c.}\right]\,. }\end{eqnarray}
That way the $\sigma_y$-contribution is suppressed, all contributions commute, ideally the state $\ket{g_{0}}$ is never populated in the interaction frame and importantly the operators appearing in the gate Hamiltonian part (the second term in\,(\ref{hmwint})) commute what allows to integrate the time evolution exactly as described in \ref{append_exact_int} (corresponding to $\hat{O}=1/2\,(3/4\,\sigma_x^{\mathrm{pm}}-1/4\,\mathds{1}_{\mathrm{pm}})$ in equation\,(\ref{exint3})). Note that, as already discussed in the double-path setup, the uncorrelated single flip interactions can be removed by a single echo $\pi$-pulse in $\sigma_y^{\mathrm{pm}}$ or $\sigma_z^{\mathrm{pm}}$. Therefore, including an echo-pulse the effective time evolution is exactly given by
\begin{equation}\fl \quad U(t_{\mathrm{gate}})=\mathrm{exp}\left(-i\,\left[ \frac{\Omega_{\mathrm{gate}}}{2}\,\sigma_x^{\mathrm{pm}}\otimes \sigma_x^{\mathrm{pm}} \right]\,t_{gate}  \right)\qquad \mathrm{with}\quad \Omega_{\mathrm{gate}}=\frac{9}{8}\,\frac{\tilde{\Omega}^2}{8\,\Delta\epsilon} \end{equation} 
if the time condition $\Delta\epsilon\,\,t_{\mathrm{gate}}=n\,(2\pi)$ ($n\in\mathds{N}$) is fulfilled. That is, for the optimal choice with respect to the gate time $n=2$ ($n=1$ cannot be realized due to the required echo pulse) this leads to $\kappa_2=\tilde{\Omega}/\Delta\epsilon=\sqrt{(4/3)^2\,\theta/\pi}$ for performing a two qubit rotation $\Omega_{\mathrm{gate}}\cdot t_{gate}=\theta$. Here it is interesting to note that the (maximal) ratio between the gate speed and the effective excited state decay rate is independent of the absolute values of the laser Rabi frequency (independent of $\kappa_1=\Omega_2/\nu=\Omega_1/\epsilon_1$ as defined in the main text) and therefore the condition $\Omega_{MW}\gg \tilde{\Omega} $ does not alter the maximal nanodiamond size limitations as depicted in figure\,\ref{b_gatesetup}. However the absolute magnitude of the gate speed ($\sim \kappa_1^2$) decreases with a decreasing microwave field, such that other limitations as the $T_2$-time of the ground state triplet states can replace the effective excited state decay rate as the limiting quantity. \par
Note also that the continuous microwave driving decouples the gate interaction from ground state decoherence, an effect that can be modelled as a fluctuating energy shift $H_{\mathrm{decoh}}=\delta(t)/2\,S_z$ and is suppressed according to equation\,(\ref{mw_subst}),
 such that the limiting quantity will be given merely by $T_1$ for a strong driving (for a more complete discussion of the decoupling method we refer to\,\cite{albrecht13}). As a final remark, the interaction frame with respect to $\sum_{j=1,2}H_{MW}^{(\mathrm{j})}$ can be implemented using the pulse sequence (see figure\,\ref{b_mw_gate}\,(c))
\begin{equation}  \mathrm{exp}\left(-i\,H_{\mathrm{int}}\,t\right)=U_{\mathrm{int}} e^{-i\,H\,t/2} U_{\mathrm{int}}^\dagger e^{-i\,H\,t/2}  \end{equation}
with $H$ given by\,(\ref{hmw1}) and $U_{int}=\exp(-i\,\pi\,S_e)$ and $S_e\in\{\sigma_x^{\mathrm{pm}}, S_y, S_z \}$ (noting that $U_{\mathrm{int}}S_xU_{\mathrm{int}}^\dagger=-S_x$ whereas $H_{int}$ is invariant under this pulse sequence).

\section{Influence of dipolar couplings on the gate interaction}\label{append_dip_coupl}
On the nanometer distance between NV centers, dipolar interactions can play a significant role and might itself provide the conditional coupling interaction at the same time disturbing the phonon induced mechanism. The optical dipolar interaction on the ground-excited state manifold can be described by\,\cite{lukin00, hettich02}
\begin{equation}  H_{\mathrm{opt,dip}}=\frac{j_{\mathrm{opt}}}{2}\,\left( \ket{e,g_{+1}}\bra{g_{+1},e}+\ket{e,g_{-1}}\bra{g_{-1}, e} +\mathrm{h.c.}\right)   \end{equation}
with the coupling constant\,\cite{hettich02, waldermann07} ($n\,k_0\,r\ll1$)
\begin{eqnarray}\eqalign{ j_{\mathrm{opt}}&=\frac{3}{2}\,\frac{\Gamma\,\xi_0}{(n\,k_0\,r)^3}\,\left( \hat{p}_1\cdot\hat{p}_2-3\,(\hat{p}_1\cdot \hat{e}_r)\,\left( \hat{p}_2\cdot\hat{e}_r \right) \right)\\
&\simeq 2\pi\cdot 52.4\,MHz\,\left(  \frac{10\,nm}{r}\right)^3\,\left( \hat{p}_1\cdot\hat{p}_2-3\,(\hat{p}_1\cdot \hat{e}_r)\,\left( \hat{p}_2\cdot\hat{e}_r \right) \right)} 
\end{eqnarray}
with $r$ the distance between the NV centers, $\Gamma$ the spontaneous decay rate ($\Gamma=15\,MHz$\,\cite{togan10, tamarat06}), $k_0=2\pi/\lambda_0$ the vacuum wavevector of the transition and $\lambda_0=637\,nm$, $n$ the refractive index ($n=2.4$), $\xi_0$ the fraction of emissions into the zero phonon line ($\xi_0\simeq 0.03$\,\cite{santori10}), $\hat{p}_i$ the normalized dipole moment direction of NV center $i$ and $\hat{e}_r$ the unit vector in the direction of the axis connecting the two NV centers. Additionally there exists a magnetic dipolar interaction within the ground state triplet manifold, that, nevertheless being orders of magnitude weaker than the optical one, can play a role for far detuned gates. It takes the form\,\cite{neumann10, albrecht13}
\begin{eqnarray} 
\eqalign{&H_{\mathrm{mag,dip}}=\frac{j_{\mathrm{mag}}}{2}\,\sigma_z^1\,\sigma_z^2 \quad \\&\mathrm{with}\quad j_{\mathrm{mag}}=2\,\left( \frac{\mu_0}{4\,\pi}\,\frac{\gamma_{el}^2\,\hbar}{r^3}\right)\,\left( \hat{p}_1\cdot\hat{p}_2-3\,(\hat{p}_1\cdot \hat{e}_r)\,\left( \hat{p}_2\cdot\hat{e}_r \right) \right) \\
&\qquad\qquad\simeq 2\pi\cdot 104\,kHz \left(\frac{10\,nm}{r}\right)^2\,\left( \hat{p}_1\cdot\hat{p}_2-3\,(\hat{p}_1\cdot \hat{e}_r)\,\left( \hat{p}_2\cdot\hat{e}_r \right) \right)}  
\end{eqnarray} 
wherein $\mu_0$ denotes the magnetic permeability, $\gamma_{el}$ the gyromagnetic ratio of the electron spin and $\sigma_z$ the Pauli z-matrix defined within\,$\{\ket{g_{+1}}, \ket{g_{-1}}  \}$.\par
\begin{figure}[htb]
\begin{centering}
\includegraphics[scale=0.5]{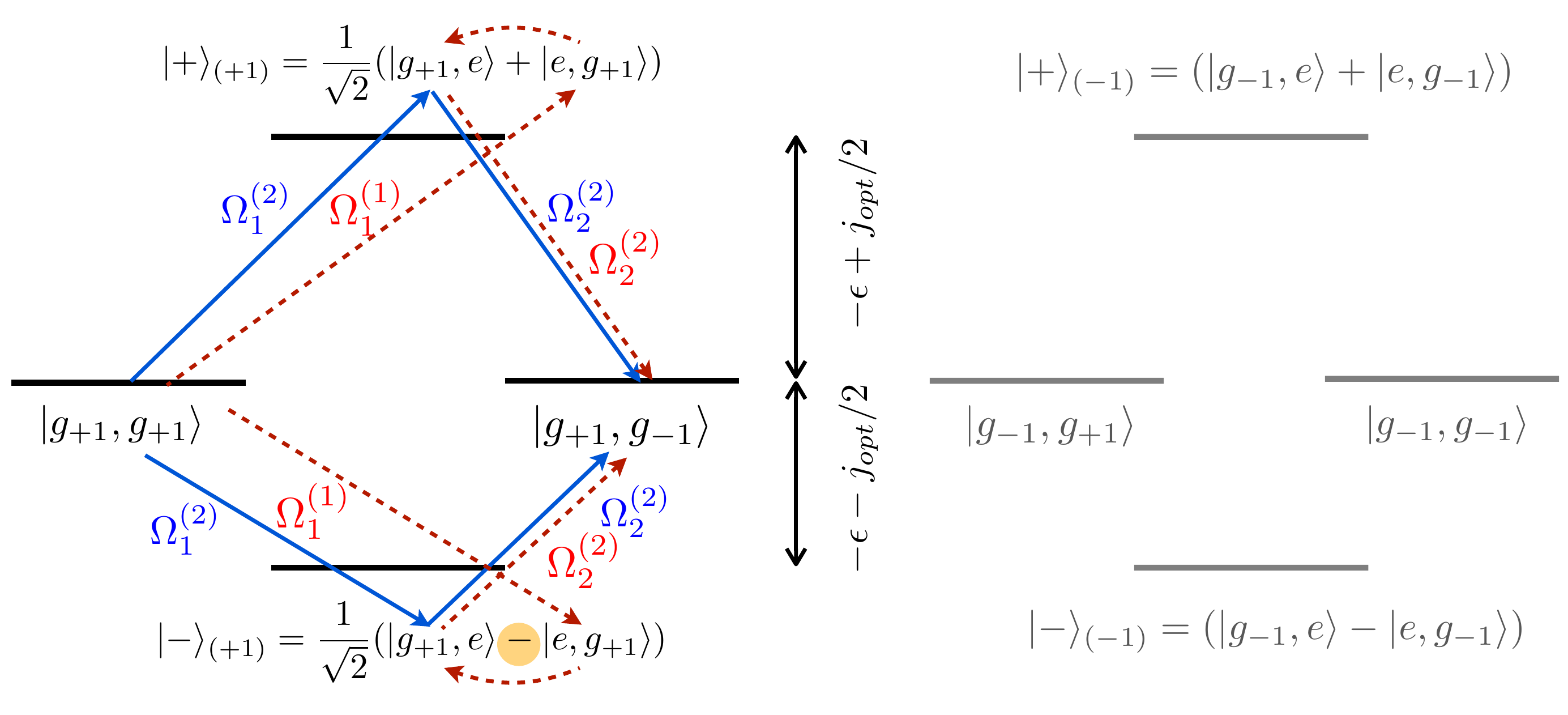}
\caption{\label{s_dipolar} \textbf{Dipolar coupling (single flip process): } Illustration of a single flip process of NV center 2 from $\ket{g_{+1}}$ to $\ket{g_{-1}}$ in the presence of optical dipolar interactions. The initial and final states are coupled to the dressed states $\ket{+}_{(+1)}$ and $\ket{-}_{(+1)}$ here illustrated for the case $|j_{\mathrm{opt}}|>|\epsilon|$ (assuming $\epsilon_1\simeq \epsilon_2\simeq \epsilon$) and $\epsilon<0$ (red detuned). Blue lines denote couplings that are independent of the dipolar interaction, whereas red dashed ones describe dipolarly induced processes.  Upper indices denote the corresponding NV center and lower ones refer to a specific coupling type. Paths related to the $\ket{-}_{(+1)}$ state cancel by interference due to the negative sign if $\Omega_1^{(1)}=\Omega_1^{(2)}$, i.e. in general if the configurations on both NV centers are identical. Analogue processes can be identified for other single flip and $\ket{g_{+1},g_{-1}}\leftrightarrow\ket{g_{-1},g_{+1}}$ conditional coupling processes.   }
\end{centering}
\end{figure}

For the derivation of the first effective Hamiltonian form\,(\ref{heff1}, \ref{heff1b}) the influence of the magnetic dipolar interaction on the excited state elimination can be neglected, and for the double path scheme the following Hamiltonian is obtained assuming identical configurations on both NV centers
\begin{eqnarray}\label{sheff1dp}  \eqalign{ H_{\mathrm{eff}}^{\mathrm{I,dp}}=&\frac{\delta^{\mathrm{dp}}}{2}\sum_k\,\sigma_x^k +\left[  i\,e^{i\,(\epsilon_1-\epsilon_2)\,t}\,a^\dagger\,\frac{1}{2}\,\left(\tilde{\Omega}\sigma_x^k+\Omega_a\,\mathds{1}\right)+\mathrm{h.c.}\right]\\
&+ \frac{\tilde{j}_{\mathrm{opt}}}{2}\,\left( \sigma_x^1\,\sigma_x^2+\sigma_y^1\,\sigma_y^2+\sigma_z^1\,\sigma_z^2 \right)+\frac{j_{\mathrm{mag}}}{2}\,\sigma_z^1\,\sigma_z^2}
\end{eqnarray}
with
\begin{eqnarray}
\eqalign{\delta^{\mathrm{dp}}&= -\frac{1}{2}\,\left(-\frac{\Omega_1^2}{(j_{\mathrm{opt}}/2)-\epsilon_1}-\frac{(1+n)\,\eta^2\,\Omega_2^2}{(j_{\mathrm{opt}}/2)-\epsilon_2} -\frac{\Omega_2^2}{(j_{\mathrm{opt}}/2)-(\nu+\epsilon_2)} \right)\\
\tilde{\Omega}&=-\frac{1}{4}\,(\eta\Omega_2)\,\Omega_1\,\left( \frac{1}{(j_{\mathrm{opt}}/2)-\epsilon_1}+\frac{1}{(j_{\mathrm{opt}})/2-\epsilon_2} \right)\\
\Omega_a&=\frac{1}{2}\,\tilde{\Omega}-\frac{1}{8}\,(\eta\Omega_2)\,\Omega_1\,\left(\frac{\epsilon_1}{(j_{\mathrm{opt}}/2)^2-\epsilon_1^2}+\frac{\epsilon_2}{(j_{\mathrm{opt}}/2)^2-\epsilon_2^2}  \right)\\
\tilde{j}_{\mathrm{opt}}&= -\frac{1}{4}\,j_{\mathrm{opt}}\,\left[  \frac{\Omega_1^2}{(j_{\mathrm{opt}}/2)^2-\epsilon_1^2}+\frac{\eta^2\,\Omega_2^2\,(1+n)}{(j_{\mathrm{opt}}/2)^2-\epsilon_2^2}\right.\\
&\left.+\frac{\eta^2\,\Omega_1^2\,\Omega_2^2\,j_{\mathrm{opt}}}{16\,\Delta\epsilon}\,\left(\frac{1}{(j_{\mathrm{opt}}/2)^2-\epsilon_1^2}+\frac{1}{(j_{\mathrm{opt}}/2)-\epsilon_2^2} \right)^2+\frac{\Omega_2^2}{(j_{\mathrm{opt}}/2)^2-\delta_2^2} \right]\,.}
 \end{eqnarray}
This effective form is valid for any magnitude of the dipolar coupling as long as $|\Omega_1|\simeq |\eta\Omega_2| \ll |(j_{\mathrm{opt}}/2)-\epsilon_1|$ and additionally $|\kappa_1^2(j/\epsilon_k)|\ll 1$ for $|j|>|\epsilon_k|$ to avoid a quasi resonant second order process to the double excited state $\ket{ee}$. Compared to the derivation without dipolar interaction\,(\ref{heff1}), the dipolar modified Hamiltonian follows essentially by replacing the detuning $\epsilon_k\rightarrow \epsilon_k-(j_{\mathrm{opt}}/2)$ what has a clear interpretation as a coupling to the dressed states of the optical dipolar interaction (see figure\,\ref{s_dipolar}). Additionally direct off-resonantly suppressed dipolar interaction terms of the order $\tilde{j}_{\mathrm{opt}}\sim j_{\mathrm{opt}}\,\kappa_1^2$ appear in the Hamiltonian together with the magnetic equivalents. Interestingly, for the case of equal couplings on both NV centers, only couplings to the $\ket{+}$ dressed state occur (except for the state independent phonon terms described by $\Omega_a$), explaining why the detuning $\epsilon_k+(j_{\mathrm{opt}}/2)$ is not a relevant quantity in the effective form\,(\ref{sheff1dp}). This is originated in the fact that $\ket{-}$ couplings cancel by interference due to the different sign of both paths as illustrated in figure\,\ref{s_dipolar}. However note that this is only valid for identical coupling configurations on both NV centers, otherwise replacements of the form (here for k=1, upper indices refer to $k$)
\begin{equation}\fl \qquad\quad \frac{\Omega_1\,\Omega_2}{j_{\mathrm{opt}}/2-\epsilon}\rightarrow \frac{\Omega^{(1)}_1\,\Omega_2^{(1)}\epsilon}{(j_{\mathrm{opt}}/2)^2-\epsilon^2}+\frac{1}{2}\frac{j_{\mathrm{opt}}}{2}\,\left(\frac{\Omega_1^{(1)}\,\Omega_2^{(2)}}{(j_{\mathrm{opt}}/2)^2-\epsilon^2}+\frac{\Omega_1^{(2)}\,\Omega_2^{(1)}}{(j_{\mathrm{opt}}/2)^2-\epsilon^2}  \right)  \end{equation}
are required in $\delta_k^{\mathrm{dp}}$, $\tilde{\Omega}_k$ and $\Omega_a$, taking into account couplings to both dressed states with the first part describing the dipolar independent and the second one the dipolar induced contributions. The terms in $\tilde{j}_{\mathrm{opt}}$ always involve terms of both $k=1$ and $k=2$ such that the replacements are obvious in that case. \par
In the absence of the magnetic dipolar term in\,(\ref{sheff1dp}) all terms commute and therefore the optical dipolar contribution just adds to the phonon induced gate interaction that can be derived from the first two terms as in\,(\ref{heff2}). A small modification of this behaviour arises from the magnetic dipolar contributions, that do not commute with the $\sigma_x$-terms in\,(\ref{sheff1dp}). However due to $j_{\mathrm{mag}}\ll (\delta^{dp}, \tilde{\Omega})$ the magnetic dipolar term contributes to\,(\ref{sheff1dp}) in the secular approximation as
\begin{equation}  H_{\mathrm{mag,dip}}\simeq \frac{j_{\mathrm{mag}}}{4}\,(\sigma_z^1\,\sigma_z^2+\sigma_y^1\,\sigma_y^2) \end{equation}
that commutes with all other contributions. Thus, the second effective Hamiltonian form analogue to\,(\ref{heff2}) is given by
\begin{eqnarray}
\eqalign{ H_{\mathrm{eff}}^{\mathrm{II,dp}}=&\left(\frac{\delta^{dp}}{2}+\frac{\Omega_a\,\tilde{\Omega}}{\Delta\epsilon}\right)\sum_k \sigma_x^k -\frac{\tilde{\Omega}^2}{2\,\Delta\epsilon} \,\sigma_x^1\,\sigma_x^2\\
&+\frac{\tilde{j}_{\mathrm{opt}}}{2}\,\left( \sigma_x^1\,\sigma_x^2+\sigma_y^1\,\sigma_y^2+\sigma_z^1\,\sigma_z^2 \right)+\frac{j_{\mathrm{mag}}}{4}\,(\sigma_z^1\,\sigma_z^2+\sigma_y^1\,\sigma_y^2)} \end{eqnarray}
wherein the single flip-contributions (the first term) can again be removed by an echo pulse and all contributions commute. The $\sigma_z^1\sigma_z^2$-contributions just correspond to a global phase in the $\mathcal{M}_1$ and $\mathcal{M}_2$-manifolds such that in total the gate frequency is given by $\Omega_{\mathrm{gate}}=-\tilde{\Omega}^2/\Delta\epsilon+2\,\tilde{j}_{\mathrm{opt}}+j_{\mathrm{mag}}/2$ for a $\mathcal{M}_1$ rotation, and $\Omega_{\mathrm{gate}}=-\tilde{\Omega}^2/\Delta\epsilon-j_{\mathrm{mag}}/2$ for a $\mathcal{M}_2$-rotation, respectively. \par
A similar analysis can be carried out for the single-path setup leading to the same replacements of the detunings by $\epsilon\rightarrow \epsilon-(j_{\mathrm{opt}}/2)$ in the uncoupled form\,(\ref{heff1b}) and direct dipolar contributions as in\,(\ref{sheff1dp}) with
\begin{equation}\fl \qquad  \tilde{j}_{\mathrm{opt}}= -\frac{1}{8}\,j_{\mathrm{opt}}\,\left[  \frac{\Omega_1^2}{(j_{\mathrm{opt}}/2)^2-\epsilon_1^2}+\frac{\eta^2\,\Omega_2^2\,(1+n)}{(j_{\mathrm{opt}}/2)^2-\epsilon_2^2}+\frac{\Omega_2^2}{(j_{\mathrm{opt}}/2)^2-(\nu+\epsilon_2)^2} \right]\,. \end{equation} 
In total the gate interaction arises from replacing $\Omega_{\mathrm{gate}}$ by $\Omega_{\mathrm{gate}}-4\,\tilde{j}_{\mathrm{opt}}$ in\,(\ref{heff2b}) beside the detuning replacements.

\section{Mode decay and excited state decay}\label{append_decay}
In here we will briefly comment on how the spontaneous decay and the coupling among vibrational modes can be incorporated into the gate formalism. Both effects can be described using a master equation approach. For the spontaneous excited state decay this takes the form
\begin{equation}\label{spdec}  \frac{\partial \rho}{\partial t}\biggr|_{\mathrm{sp.\,dec}}=\frac{\Gamma}{2}\,\left( 2\,\tilde{\sigma}_-\rho\tilde{\sigma}_+-\left\{\rho, \sigma_+\sigma_- \right\} \right)   \end{equation}
with $\sigma_+=\ket{e}\left(\bra{g_{+1}}+\bra{g_{-1}}\right)$ and $\sigma_-=\left(\sigma_+\right)^\dagger$, $\Gamma$ the excited state decay rate, $\{a,b  \}=ab+ba$ denoting the anti-commutator and $\tilde{\sigma}_{\pm}=\sigma_\pm\,e^{\pm i\,\eta\left(a^\dagger+a\right)}$. The phonon mode relaxation can be modelled as\,\cite{wilson04}
\begin{equation}\label{phonrelax}\fl \qquad  \frac{\partial \rho}{\partial t} \biggr|_{\mathrm{mode relax.}} =\frac{\nu}{Q}\,(n_{th}+1)\,\left[ a\,\rho\,a^\dagger-\frac{1}{2}\,\left\{a^\dagger\,a,\, \rho \right\} \right]+\frac{\nu}{Q}\,n_{th}\left[  a^\dagger\,\rho\,a-\frac{1}{2}\,\left\{a\,a^\dagger,\, \rho \right\} \right]  \end{equation}
with $\nu$ the mode frequency as defined in the main text, $Q$ the corresponding mode quality factor and $n_{th}$ the thermal phonon population. Both of those equations hold on the level of the original Hamiltonian after the Schrieffer-Wolff transformation, i.e. in the same frame as Hamiltonian\,(\ref{hrotframe}).\par
The elimination of the excited state manifold, i.e. calculating an effective Hamiltonian as in \,(\ref{heff1}, \ref{heff1b}), corresponds to a unitary transformation $T$\,\cite{tannoudji04} such that $H_{\mathrm{eff}}^{\mathrm{I}}=\mathcal{P}_gT\,H\,T^\dagger \mathcal{P}_g$ with $\mathcal{P}_g$ the projector on the ground state manifold and $H(t)$ the original Hamiltonian as e.g. given by\,(\ref{hrotframe}). Up to second order the transformation is given by\,\cite{tannoudji04, james07} (for the case of a Hamiltonian that is purely off-diagonal, e.g. only includes terms that couple the ground to the excited state)
\begin{equation} T=\mathds{1}+i\,S \qquad \mathrm{with} \qquad S=\int^t H(t')\,\mathrm{d}t'  \end{equation}
with $H(t')$ the ground-excited state coupling Hamiltonian ($H_k^{sp}$\,(\ref{hrotframe}) for the single path setup or $H_k^{dp}$ for the double-path equivalent),
and the effective Hamiltonian follows as
\begin{equation}  H_{\mathrm{eff}}^{\mathrm{I}}(t)=\frac{1}{2}\,i\,\,\mathcal{P}_g \, \left[S,H(t) \right]\,\mathcal{P}_g\,. \end{equation}
An arbitrary operator $\hat{O}$ in that effective frame can therefore be calculated by
\begin{equation}\label{eff_op} \hat{O}_{\mathrm{eff}}^{\mathrm{I}}=\mathcal{P}_g\,\left( \hat{O}+i\,\left[S,\hat{O}  \right]-\frac{1}{2}\,\left[S,  \left[ S,\hat{O}  \right] \right]  \right)\,\mathcal{P}_g  \end{equation}
which for the specific case of the ladder operators $\sigma_-$ and $a$ results in
\begin{equation} \label{eff_sigmas}  \sigma_-^{\mathrm{eff,I}}=-i\,\sigma_-\,S, \qquad a_{\mathrm{eff,I}}=a-\frac{1}{2}\,\left[S,  \left[ S,\hat{a}  \right] \right]   \end{equation}
and $\sigma_+^{\mathrm{eff,I}}=\left(\sigma_-^{\mathrm{eff,I}}\right)^\dagger$ and $a^{\dagger}_{\mathrm{eff,I}}=\left( a_{\mathrm{eff,I}} \right)^\dagger$. Therefore the evolution in the first effective frame is correctly described by replacing the operators appearing in the master equations\,(\ref{spdec}) and (\ref{phonrelax}) with the effective ones as defined in\,(\ref{eff_op},\ref{eff_sigmas}). Additionally one has to account for the ground state dephasing by either a stochastic approach or a master equation approach in the Markovian limit\,\cite{albrecht13}, unchanged by the unitary transformation for the first effective form, as $T$ is diagonal by construction within the ground state manifold.     \par
To give an explicit example we will consider the double-path setup and assume equal coupling configurations $\eta_1=\eta_2=\eta$, which results in (with $\sigma_x$, $\mathds{1}$ defined within the ground state manifold, e.g. $\sigma_x=\ket{g_{+1}}\bra{g_{-1}}+\mathrm{h.c.}$)
\begin{eqnarray}\label{seff_dp}
\fl \qquad\quad \sigma_{-}^{\mathrm{eff,I}}&=\left(\mathds{1}+\sigma_x \right)\,\left( \frac{\Omega_1}{2\,\epsilon_1}\,e^{-i\,\epsilon_1\,t}+\frac{\Omega_2}{2(\epsilon_2+\nu)}\,e^{-i(\epsilon_2+\nu)\,t}+i\,a^\dagger\,\frac{\eta\Omega_2}{2\epsilon_2}\,e^{-i\,\epsilon_2\,t} \right) \\
\fl \qquad\quad a_{\mathrm{eff,I}}&= a-\frac{1}{2}\left(\sigma_x +\mathds{1} \right)\,\frac{\eta\Omega_2}{2\epsilon_2}\,\left(-i\,\frac{\Omega_1}{2\epsilon_1}\,e^{i\,\Delta\epsilon\,t}-\frac{\eta\Omega_2}{2\epsilon_2}\,a-i\frac{\Omega_2}{2(\epsilon_2+\nu)}\,e^{i\nu t}  \right)\,.
 \end{eqnarray}
Out of\,(\ref{seff_dp}) it is obvious that the decay rate $\Gamma$ appearing in\,(\ref{spdec}) can be replaced by an effective one of the order $\Gamma_{\mathrm{eff,I}}\sim \kappa_1^2$ as expected by the fact that this corresponds to the probability of populating the excited state.

\clearpage
\section*{References}
\bibliography{phonon_coupl}

\providecommand{\newblock}{}
\begin{thebibliography}{10}
\expandafter\ifx\csname url\endcsname\relax
  \def\url#1{{\tt #1}}\fi
\expandafter\ifx\csname urlprefix\endcsname\relax\def\urlprefix{URL }\fi
\providecommand{\eprint}[2][]{\url{#2}}

\bibitem{jelezko04}
Jelezko F, Gaebel T, Popa I, Gruber A and Wrachtrup J 2004 {\em Phys. Rev.
  Lett.\/} {\bf 92} 076401

\bibitem{balasubramanian09}
Balasubramanian G, Neumann P, Twitchen D, Markham M, Kolesov R, Mizuochi N,
  Isoya J, Achard J, Beck J, Tissler J {\em et~al.\/} 2009 {\em Nature
  Mater.\/} {\bf 8} 383--387

\bibitem{gaebel06}
Gaebel T, Domhan M, Popa I, Wittmann C, Neumann P, Jelezko F, Rabeau J,
  Stavrias N, Greentree A, Prawer S {\em et~al.\/} 2006 {\em Nature Phys.\/}
  {\bf 2} 408--413

\bibitem{vanderSar12}
van~der Sar T, Wang Z, Blok M, Bernien H, Taminiau T, Toyli D, Lidar D,
  Awschalom D, Hanson R and Dobrovitski V 2012 {\em Nature\/} {\bf 484} 82--86

\bibitem{dutt07}
Dutt M, Childress L, Jiang L, Togan E, Maze J, Jelezko F, Zibrov A, Hemmer P
  and Lukin M 2007 {\em Science\/} {\bf 316} 1312--1316

\bibitem{neumann08}
Neumann P, Mizuochi N, Rempp F, Hemmer P, Watanabe H, Yamasaki S, Jacques V,
  Gaebel T, Jelezko F and Wrachtrup J 2008 {\em Science\/} {\bf 320} 1326--1329

\bibitem{dolde13}
Dolde F, Jakobi I, Naydenov B, Zhao N, Pezzagna S, Trautmann C, Meijer J,
  Neumann P, Jelezko F and Wrachtrup J 2013 {\em Nature Phys.\/} {\bf 9}
  139--143

\bibitem{bernien12}
Bernien H, Hensen B, Pfaff W, Koolstra G, Blok M~S, Robledo L, Taminiau T~H,
  Markham M, Twitchen D~J, Childress L and Hanson R 2013 {\em Nature\/} {\bf
  497} 86--90

\bibitem{albrecht13}
Albrecht A, Retzker A, Koplovitz G, Jelezko F, Yochelis S, Porath D, Nevo Y,
  Shoseyov O, Paltiel Y and Plenio M 2013 {\em arXiv:1301.1871\/}

\bibitem{bermudez11}
Bermudez A, Jelezko F, Plenio M~B and Retzker A 2011 {\em Phys. Rev. Lett.\/}
  {\bf 107}

\bibitem{wolters10}
Wolters J, Schell A, Kewes G, Nusse N, Schoengen M, Doscher H, Hannappel T,
  Lochel B, Barth M and Benson O 2010 {\em Appl. Phys. Lett.\/} {\bf 97}
  141108--141108

\bibitem{riedrich11}
Riedrich-M{\"o}ller J, Kipfstuhl L, Hepp C, Neu E, Pauly C, M{\"u}cklich F,
  Baur A, Wandt M, Wolff S, Fischer M {\em et~al.\/} 2011 {\em Nature
  Nanotech.\/} {\bf 7} 69--74

\bibitem{faraon12}
Faraon A, Santori C, Huang Z, Acosta V and Beausoleil R 2012 {\em Phys. Rev.
  Lett.\/} {\bf 109} 33604

\bibitem{hausmann12}
Hausmann B, Shields B, Quan Q, Maletinsky P, McCutcheon M, Choy J, Babinec T,
  Kubanek A, Yacoby A, Lukin M {\em et~al.\/} 2012 {\em Nano Lett.\/} {\bf 12}
  1578--1582

\bibitem{leibfried2003}
Leibfried D, Blatt R, Monroe C and Wineland D 2003 {\em Rev. Mod. Phys.\/} {\bf
  75} 281

\bibitem{soykal11}
Soykal O~O, Ruskov R and Tahan C 2011 {\em Phys. Rev. Lett.\/} {\bf 107}(23)
  235502

\bibitem{kolkowitz12}
Kolkowitz S, Jayich A, Unterreithmeier Q, Bennett S, Rabl P, Harris J and Lukin
  M 2012 {\em Science\/} {\bf 335} 1603--1606

\bibitem{arcizet11}
Arcizet O, Jacques V, Siria A, Poncharal P, Vincent P and Seidelin S 2011 {\em
  Nature Phys.\/} {\bf 7} 879--883

\bibitem{hong12}
Hong S, Grinolds M~S, Maletinsky P, Walsworth R~L, Lukin M~D and Yacoby A 2012
  {\em Nano Lett.\/} {\bf 12} 3920--3924

\bibitem{rabl10}
Rabl P, Kolkowitz S, Koppens F, Harris J, Zoller P and Lukin M 2010 {\em Nature
  Phys.\/} {\bf 6} 602--608

\bibitem{tisler09}
Tisler J, Balasubramanian G, Naydenov B, Kolesov R, Grotz B, Reuter R, Boudou
  J~P, Curmi P~A, Sennour M, Thorel A {\em et~al.\/} 2009 {\em ACS Nano\/} {\bf
  3} 1959--1965

\bibitem{maze11}
Maze J, Gali A, Togan E, Chu Y, Trifonov A, Kaxiras E and Lukin M 2011 {\em New
  J. Phys.\/} {\bf 13} 025025

\bibitem{doherty11}
Doherty M, Manson N, Delaney P and Hollenberg L 2011 {\em New J. Phys.\/} {\bf
  13} 025019

\bibitem{ashcroft05book}
Ashcroft N and Mermin N 2006 {\em Solid State Physics\/} (Brooks/Coole Thomson
  Learning)

\bibitem{yu99book}
Yu P and Cardona M 1999 {\em Fundamentals of semiconductors: physics and
  materials properties\/} (Springer Berlin)

\bibitem{togan11}
Togan E, Chu Y, Imamoglu A and Lukin M 2011 {\em Nature\/} {\bf 478} 497--501

\bibitem{wilson04}
Wilson-Rae I, Zoller P and Imamo{\=g}lu A 2004 {\em Phys. Rev. Lett.\/} {\bf
  92} 75507

\bibitem{mintert01}
Mintert F and Wunderlich C 2001 {\em Phys. Rev. Lett.\/} {\bf 87} 257904

\bibitem{cirac93}
Cirac J, Blatt R, Parkins A and Zoller P 1993 {\em Phys. Rev. A\/} {\bf 48}
  2169

\bibitem{sorensen99}
S{\o}rensen A and M{\o}lmer K 1999 {\em Phys. Rev. Lett.\/} {\bf 82} 1971--1974

\bibitem{bermudez12}
Bermudez A, Schmidt P~O, Plenio M~B and Retzker A 2012 {\em Phys. Rev. A\/}
  {\bf 85} 040302

\bibitem{lemmer13}
Lemmer A, Bermudez A and Plenio M~B 2013 {\em arXiv:1303.5770\/}

\bibitem{roos08}
Roos C 2008 {\em New J. Phys.\/} {\bf 10} 013002

\bibitem{togan10}
Togan E, Chu Y, Trifonov A, Jiang L, Maze J, Childress L, Dutt M, Soerensen A
  and Hemmer P 2010 {\em Nature\/} {\bf 466} 730--734

\bibitem{greffet11}
Greffet J~J, Hugonin J~P, Besbes M, Lai N, Treussart F and Roch J~F 2011 {\em
  arXiv:1107.0502\/}

\bibitem{beveratos01}
Beveratos A, Brouri R, Gacoin T, Poizat J~P and Grangier P 2001 {\em Phys. Rev.
  A\/} {\bf 64} 061802

\bibitem{neumann09}
Neumann P, Kolesov R, Jacques V, Beck J, Tisler J, Batalov A, Rogers L, Manson
  N, Balasubramanian G, Jelezko F {\em et~al.\/} 2009 {\em New J. Phys.\/} {\bf
  11} 013017

\bibitem{lukin00}
Lukin M and Hemmer P 2000 {\em Phys. Rev. Lett.\/} {\bf 84} 2818--2821

\bibitem{neukirch13}
Neukirch L~P, Gieseler J, Quidant R, Novotny L and Vamivakas A~N 2013 {\em
  arXiv:1305.1515\/}

\bibitem{robledo11}
Robledo L, Childress L, Bernien H, Hensen B, Alkemade P~F and Hanson R 2011
  {\em Nature\/} {\bf 477} 574--578

\bibitem{audenaert06}
Audenaert K~M~R and Plenio M~B 2006 {\em New J. Phys.\/} {\bf 8} 266

\bibitem{wunderlich09}
Wunderlich H and Plenio M~B 2009 {\em J. Mod. Opt.\/} {\bf 56} 2100--2105

\bibitem{bennett13}
Bennett S~D, Yao N~Y, Otterbach J, Zoller P, Rabl P and Lukin M~D 2013 {\em
  Phys. Rev. Lett.\/} {\bf 110} 156402

\bibitem{lamb1882}
Lamb H 1882 {\em Proc. Math. Soc. London\/} {\bf 13} 189--212

\bibitem{mcskimin72}
McSkimin H~J and Andreatch P 1972 {\em Journal of Applied Physics\/} {\bf 43}
  2944--2948

\bibitem{eringen75}
Eringen A~C and {\c{S}}uhubi E~S 1975 {\em Elastodynamics\/} vol~2 (Academic
  Press New York)

\bibitem{tamura82}
Tamura A, Higeta K and Ichinokawa T 1982 {\em Journal of Physics C: Solid State
  Physics\/} {\bf 15} 4975

\bibitem{takagahara96}
Takagahara T 1996 {\em Journal of Luminescence\/} {\bf 70} 129--143

\bibitem{hettich02}
Hettich C, Schmitt C, Zitzmann J, K{\"u}hn S, Gerhardt I and Sandoghdar V 2002
  {\em Science\/} {\bf 298} 385--389

\bibitem{waldermann07}
Waldermann F, Olivero P, Nunn J, Surmacz K, Wang Z, Jaksch D, Taylor R,
  Walmsley I, Draganski M, Reichart P {\em et~al.\/} 2007 {\em Diam. Relat.
  Mater.\/} {\bf 16} 1887--1895

\bibitem{tamarat06}
Tamarat P, Gaebel T, Rabeau J, Khan M, Greentree A, Wilson H, Hollenberg L,
  Prawer S, Hemmer P, Jelezko F {\em et~al.\/} 2006 {\em Phys. Rev. Lett.\/}
  {\bf 97} 83002

\bibitem{santori10}
Santori C, Barclay P, Fu K~C, Beausoleil R, Spillane S and Fisch M 2010 {\em
  Nanotechnology\/} {\bf 21} 274008

\bibitem{neumann10}
Neumann P, Kolesov R, Naydenov B, Beck J, Rempp F, Steiner M, Jacques V,
  Balasubramanian G, Markham M, Twitchen D {\em et~al.\/} 2010 {\em Nature
  Phys.\/} {\bf 6} 249--253

\bibitem{tannoudji04}
Cohen-Tannoudji C, Dupont-Roc J and Grynberg G 2004 {\em Atom-Photon
  Interactions\/} (Wiley-VCH)

\bibitem{james07}
James D and Jerke J 2007 {\em Can. J. Phys.\/} {\bf 85} 625--632

\end{thebibliography}

\end{document}